\documentclass[]{spie}  

\usepackage{amsmath,amsfonts,amssymb}
\usepackage{graphicx}
\usepackage{subcaption}
\usepackage{url}
\usepackage[colorlinks=true, allcolors=blue]{hyperref}

\title{Design and characterization of the POLARBEAR-2b and POLARBEAR-2c cosmic microwave background cryogenic receivers}

\author[a]{L. Howe}
\author[a]{C. Tsai}
\author[a]{L. Lowry}
\author[a]{K. Arnold}
\author[b]{G. Coppi}
\author[c]{J. Groh}
\author[d]{X. Guo}
\author[a]{B. Keating}
\author[c]{A. Lee}
\author[e]{A. J. May}
\author[e]{L. Piccirillo}
\author[a]{N. Stebor}
\author[a]{G. Teply}
\affil[a]{Center for Astrophysics and Space Science, University of California, San Diego, CA, USA}
\affil[b]{Department of Physics, University of Pennsylvania, Philadelphia, PA, USA}
\affil[c]{Department of Physics, University of California, Berkeley, CA, USA}
\affil[d]{Department of Physics, The Chinese University of Hong Kong, Shatin, Hong Kong, PRC}
\affil[e]{Jodrell Bank Centre for Astrophysics, School of Physics and Astronomy, University of Manchester, Manchester, UK}

\authorinfo{lahowe@ucsd.edu}

\begin{document} 
\maketitle

\begin{abstract}
The POLARBEAR-2/Simons Array Cosmic Microwave Background (CMB) polarization experiment is an upgrade and expansion of the existing POLARBEAR-1 (PB-1) experiment, located in the Atacama desert in Chile. Along with the CMB temperature and $E$-mode polarization anisotropies, PB-1 and the Simons Array study the CMB $B$-mode polarization anisotropies produced at large angular scales by inflationary gravitational waves, and at small angular scales by gravitational lensing. These measurements provide constraints on various cosmological and particle physics parameters, such as the tensor-to-scalar ratio $r$, and the sum of the neutrino masses. The Simons Array consists of three 3.5~m diameter telescopes with upgraded POLARBEAR-2 (PB-2) cryogenic receivers, named PB-2a, -2b, and -2c. PB-2a and -2b will observe the CMB over multiple bands centered at 95~GHz and 150 GHz, while PB-2c will observe at 220~GHz and 270~GHz, which will enable enhanced foreground separation and de-lensing. Each Simons Array receiver consists of two cryostats which share the same vacuum space: an optics tube containing the cold reimaging lenses and Lyot stop, infrared-blocking filters, and cryogenic half-wave plate; and a backend which contains the focal plane detector array, cold readout components, and millikelvin refrigerator. Each PB-2 focal plane array is comprised of 7,588 dual-polarization, multi-chroic, lenslet- and antenna-coupled, Transition Edge Sensor (TES) bolometers which are cooled to 250~mK and read out using Superconducting Quantum Interference Devices (SQUIDs) through a digital frequency division multiplexing scheme with a multiplexing factor of 40. In this work we describe progress towards commissioning the PB-2b and -2c receivers including cryogenic design, characterization, and performance of both the PB-2b and -2c backend cryostats. 
\end{abstract}


\section{INTRODUCTION}

\begin{figure}[t]
  \centering
  \includegraphics[width=\textwidth]{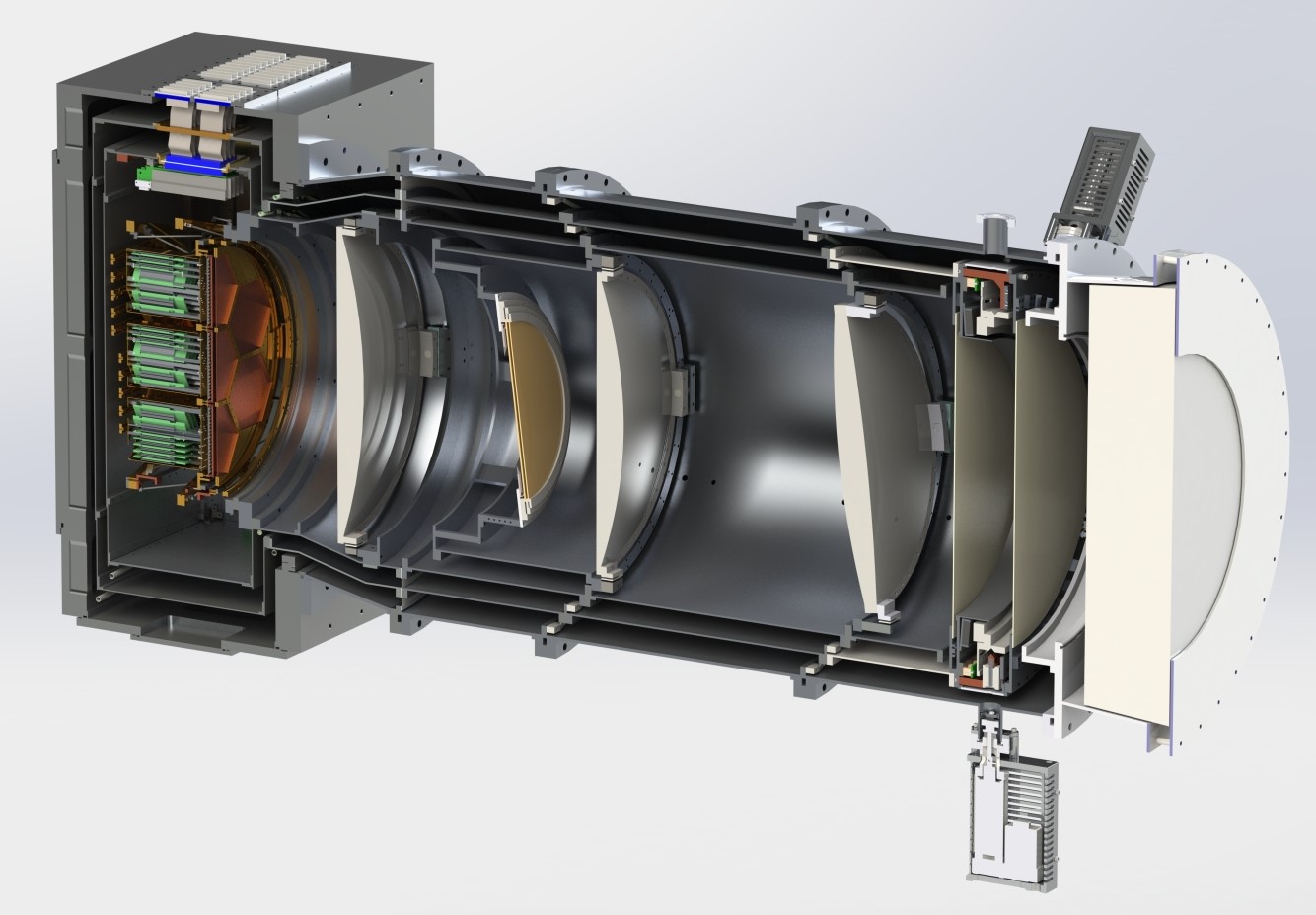}
  \vspace*{2mm}
  \caption{Section view of the PB-2b receiver CAD model. The optics tube + backend are 2~m long and the diameter of the optics tube 300~K (outermost) shell is 0.75~m. The FPT and detector modules (hexagonal structures) are located on the far left inside the rectangular backend. The SQUID pre-amplifiers and wiring harnesses are located just above the focal plane. The optics tube consists of three reimaging lenses (gray, convex circular structures), a cold aperture (yellow circular structure), a vacuum window (thick white, circular structure on far right), IR absorbing filters (thin grey flat curcilar structures) and a cryogenic half waveplate (between vacuum window and first lens). The protruding elements near the vacuum window are mechanical grippers for the half waveplate sapphire (not shown). Cryogenic refrigerators are not shown.}
  \label{fig:receiver_cad}
\end{figure}

The current accepted model of our Universe is built on precision measurements of the temperature and polarization anisotropies on the cosmic microwave background (CMB) that have been made in the last 30 years. These measurements have revealed that our Universe is both flat and currently Dark Energy-dominated, however the exact physics governing evolution during the earliest times remains a mystery. Measurement of the polarization properties of the CMB may shed light on the quantum nature of the primordial Universe, and constrain other fundamental parameters responsible for the creation of the large scale structure observed today.

It is commonly held that in order to create our Universe, which is both uniform and isotropic on scales larger than the Hubble time, there must have been a period of exponential expansion driven by a scalar field immediately following the Big Bang. This period of superluminal expansion is called \emph{cosmic inflation} and to date there is no direct evidence that this phenomenon occurred. Cosmic inflation is unique in that it would generate a stochastic background of gravitational waves at large (degree) scales whose amplitudes are proportional to the energy scale of inflation. Recent results in studies of CMB anisotropies \cite{ade2014detection, mortonson2014joint} have revealed that $B$-mode measurements are critically dependent on foreground subtraction and gravitational de-lensing, especially in efforts to measure the temperature-to-scalar ratio $r$. Current bounds set $r < 0.09$, indicating the energy scale of inflation is as high as $\sim 10^{16}$~GeV. Detection of inflationary gravitational waves could provide windows into ultra-high energy physics, including grand unified theories.

The Simons Array \cite{arnold2014simons, stebor2016simons} is a next-generation CMB polarization experiment which upgrades and expands the previous generation POLARBEAR-1 (PB-1) experiment \cite{kermish2012polarbear} to an array of three POLARBEAR-2 (PB-2) receivers -- PB-2a, -2b, and -2c -- mounted on three 3.5~m style off-axis Gregorian-Dragone telescopes, each with larger cryogenic receivers and numbers of detectors at multiple observing frequencies. Expanding the number of detectors and observing frequencies -- 95~GHz and 150~GHz for PB-2a and -2b, and 220~GHz and 270~GHz for -2c -- increases sensitivity and enables precise foreground subtraction and gravitational de-lensing, which are imperative to detect the very low amplitude inflationary $B$-mode signal. After performing cross-correlations with Planck, the Simons Array will place an upper bound of $r < 4 \times 10^{-3}$ (95\% CL).

Both Big Bang Nucleosynthesis and the standard model of particle physics contain neutrinos as critical components but very little is known about their masses. Additionally, Big Bang relic neutrinos are believed to be responsible for the formation of the large-scale gravitational structure in the Universe. By combining measurements of the CMB polarization with the DESI Baryon Acoustic Oscillation (BAO) experiment \cite{levi2013desi} the Simons Array will constrain the sum of the neutrino masses to $\sum m_\nu \leq 40$~meV (95\% CL) and determine if the mass hierarchy is inverted.

\section{The POLARBEAR-2 Cryogenic Receiver}

\begin{figure}[t]
      \centering
      \includegraphics[width = \textwidth]{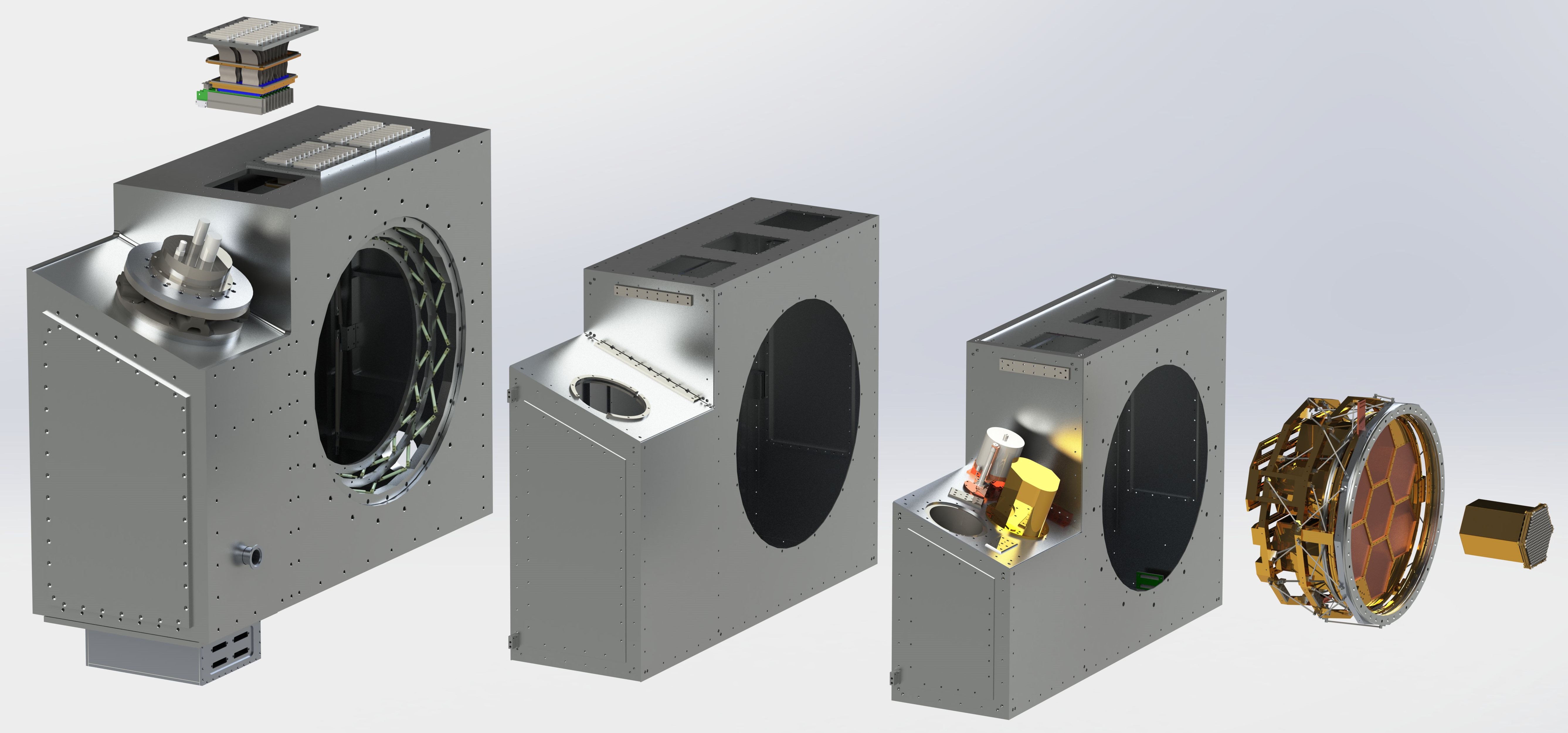}
      \vspace*{2mm}
      \caption{Exploded view of the PB-2b backend showing (from left to right) the 300~K shell, 50~K shell, 4~K shell, focal plane tower, and one detector module. Also shown exploded from the 300~K shell is one wiring harness. The pulse tube cryocooler is mounted to an anti-vibration bellows on the 300~K shell and the millikelvin refrigerators are mounted to the 4~K shell mainplate (angled portion).}
      \label{fig:backend_exploded}
\end{figure}

Increasing the number of detectors per receiver generally requires increasing the size of the cryostat. While the PB-1 receiver consists of a single cryostat for the focal plane, SQUID pre-amplifiers, re-imaging optics, and IR filters, each of the Simons Array's PB-2 receivers are constructed from two separate cryostats: the backend and the optics tube. These are fabricated and cryogenically validated separately before integration. A CAD drawing of the full PB-2b receiver is shown in Figure~\ref{fig:receiver_cad}.

Both the backend and the optics tube consist of a 300~K vacuum shell, a 50~K shell, and a 4~K shell, their namesake being derived from their approximate operating temperatures. For both cryostats the 50~K stage is used to intercept radiative loading from room temperature. The backend 4~K shell is used to cool the SQUID pre-amplifiers and provide a stage for condensation of helium-4 (He-4), which is essential for reaching millikelvin temperatures, and the optics tube 4~K shell is used to cool the reimaging lenses and cold aperture to reduce optical loading of the TES bolometers (enhances sensitivity). Cooling of the 50~K and 4~K shells in each cryostat is achieved with two PT415 two-stage mechanical pulse tube cryocoolers (PTCs) from Cryomech Inc.\footnote{\url{http://www.cryomech.com/cryorefrigerators/pulse-tube/pt415/}} (one each for the backend and optics tube). In addition to the PTC and the 300~K, 50~K, and 4~K shells, the PB-2 backend cryostat also contains the detector modules, focal plane tower (FPT), cold readout components, SQUID pre-amplifiers, cryogenic wiring for detector readout and housekeeping, and millikelvin refrigerator. An exploded view of the PB-2b backend is shown in Figure~\ref{fig:backend_exploded}. The main function of the backend is to create a $\sim$250~mK temperature stage where the detector thermal carrier noise is subdominant to the CMB photon noise. It is also important to create a $\sim$4~K SQUID stage but this is achieved as a by-product of the cryogenics required to reach millikelvin temperatures (He-4 condensation in the millikelvin refrigerator).

\begin{figure}[t]
	\centering
    \includegraphics[width=.6\textwidth]{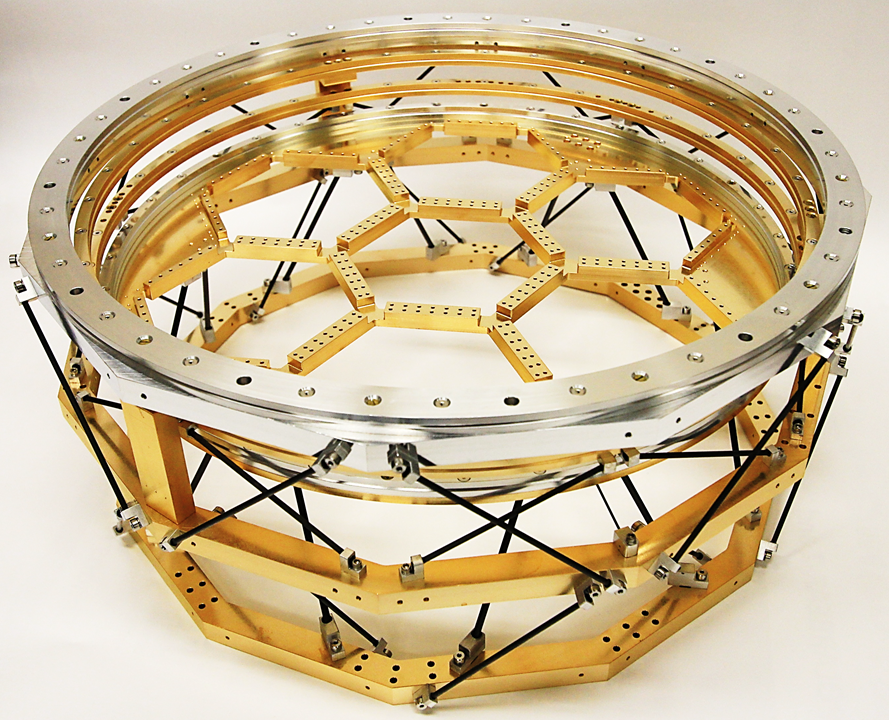}
    \vspace*{2mm}
    \caption{Photo of the PB-2b FPT which is designed around the He-10 gas light fridge and consists of four thermal stages: 4~K (top silver dodecagonal ring), 1~K (middle gold dodecagonal ring), 350~mK (bottom dodecagonal ring), and 250~mK bolometer stage (circular plate with hexagonal cutouts). The circular rings are the RF shield thermal intercepts. Not shown is the 350~mK MMF stage which is located directly above the detector modules.}
    \label{fig:fpt}
\end{figure}

The FPT, shown in Figure~\ref{fig:fpt}, is a multi-stage thermal isolation structure which is thermally anchored to each substage of the millikelvin fridge and provides thermal intercepts for the detector wiring and a radio frequency (RF) shield at 4~K, 1~K, 350~mK, and 250~mK. Additionally, there is a 350~mK stage where the final IR bandpass metal mesh filters (MMFs) are located \cite{ade2006review}. Thermal isolation of each stage is provided by carbon fiber rods which are epoxied to aluminum 6061  (Al6061) alloy feet using Stycast 2850ft \footnote{\url{http://na.henkel-adhesives.com/product-search-1554.htm?nodeid=8802688008193}}. The 1~K ring and 350~mK rings are made of Al6061 as opposed to copper to minimize weight and thermal mass while maintaining structural strength. The bolometer stage (250~mK) is made from copper 101 (C101) alloy. The majority of all metal parts, excluding the Al6061 carbon fiber feet, is plated with gold to a thickness of 1.27~$\mu$m to reduce emissivity and increase the contact conductance of the heatstraps anchoring each FPT stage to its corresponding millikelvin fridge substage.

Design of the FPT was motivated not only thermally based on results from detailed testing of the PB-2b millikelvin refrigerator (Section~\ref{sec:mK_fridge_testing}), but also from a desire to avoid coupling vibrations to the cold stages which can cause microphonic heating during observations. The FPT was designed to be a high-pass filter with its cutoff above that of the backend, which acts as a low-pass filter. This  minimizes the vibrational coupling to the FPT millikelvin stages. Lab testing has shown the FPT does behave as a high-pass filter with its cutoff frequency of $\sim$80~Hz, while the backend does not perform exactly as a low-pass filter. Rather, the backend has demonstrated broadband vibration rejection excepting 120~Hz (along the receiver z-axis), and 75~Hz (x-axis), although the latter is suspected to be a mode excited in the receiver mounting cart. Microphonic heating has not been observed in laboratory testing near these frequencies using vibration amplitudes expeced during observation.

\subsection{50~K and 4~K Refrigeration}
\begin{figure}[t]
	\centering
	\includegraphics[width = .5\textwidth]{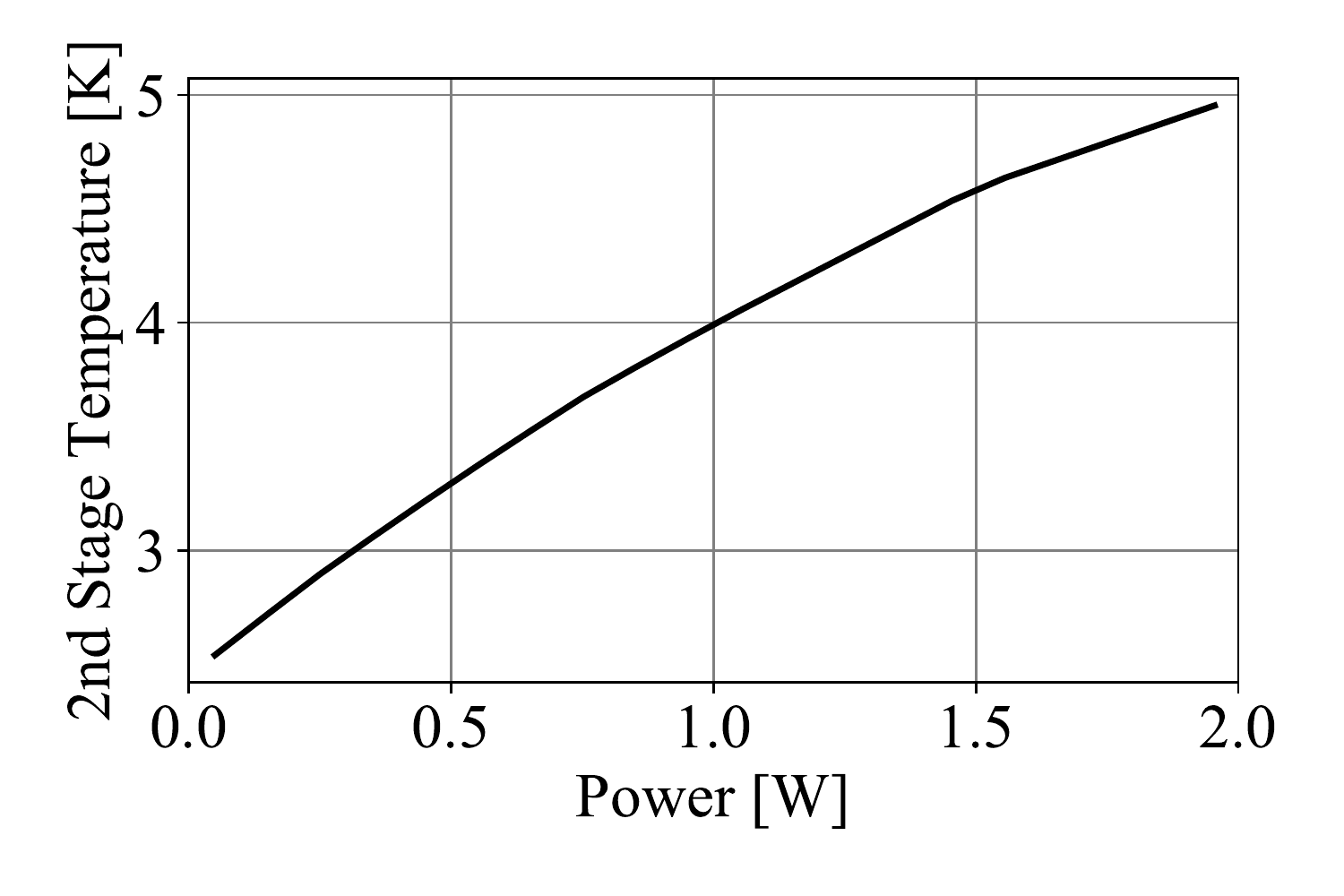}
	\caption{Load curve of the PB-2b backend PTC second stage with care taken to fully minimize conduction parasitics and radiative loads. The base temperature (0~W applied power) is 2.54~K. All increases in temperature above this zero load configuration are due to increased loading to the coldhead and can be measured absolutely using this curve.}
	\label{fig:bare_pb2b_pt415_load_curve}
\end{figure}

For all cryocoolers there is a relation between the maximum heat load that they can dissipate and their operating temperature, known as a \emph{load curve}, or in a two-parameter space, a \emph{capacity map}. The nominal cooling power of a PT415 is 50~W at 50~K on the first stage coldhead, and 1.5~W at 4~K on the second stage coldhead. Heat loads above (below) these operating points results in higher (lower) system temperatures as can be seen in the load curve of the PB-2b backend PT415 second stage shown in Figure~\ref{fig:bare_pb2b_pt415_load_curve}. A measurement of the base temperatures of the PTC stages allows for precise determination of the loading present. Measurement of the load curve of a PT415 first stage is more difficult but less variable from unit-to-unit, and has been done to high temperatures \cite{green2015cooling}. Generally, minimizing loading between stages is crucial for basic functionality of the cryostat, while precise optimization of the cryogenics is critically important for creating a high duty cycle receiver with minimal complications in readout and biasing of the detectors. The lowest temperature reached by each stage of the PTC is referred to as the \emph{base temperature}, which is a balance of the PTC cooling capacity and the parasitic loading. For CMB cryogenic receivers the most important base temperature is typically the second stage (for PB-2b  and -2c this is the 4~K shell), where operation near 3~K is preferred. The primary driver for this target is the He-4 condensation efficiency in the millikelvin refrigerator, which will be discussed in later sections.

\begin{figure}[t]
	\centering
	\includegraphics[width = .8\textwidth]{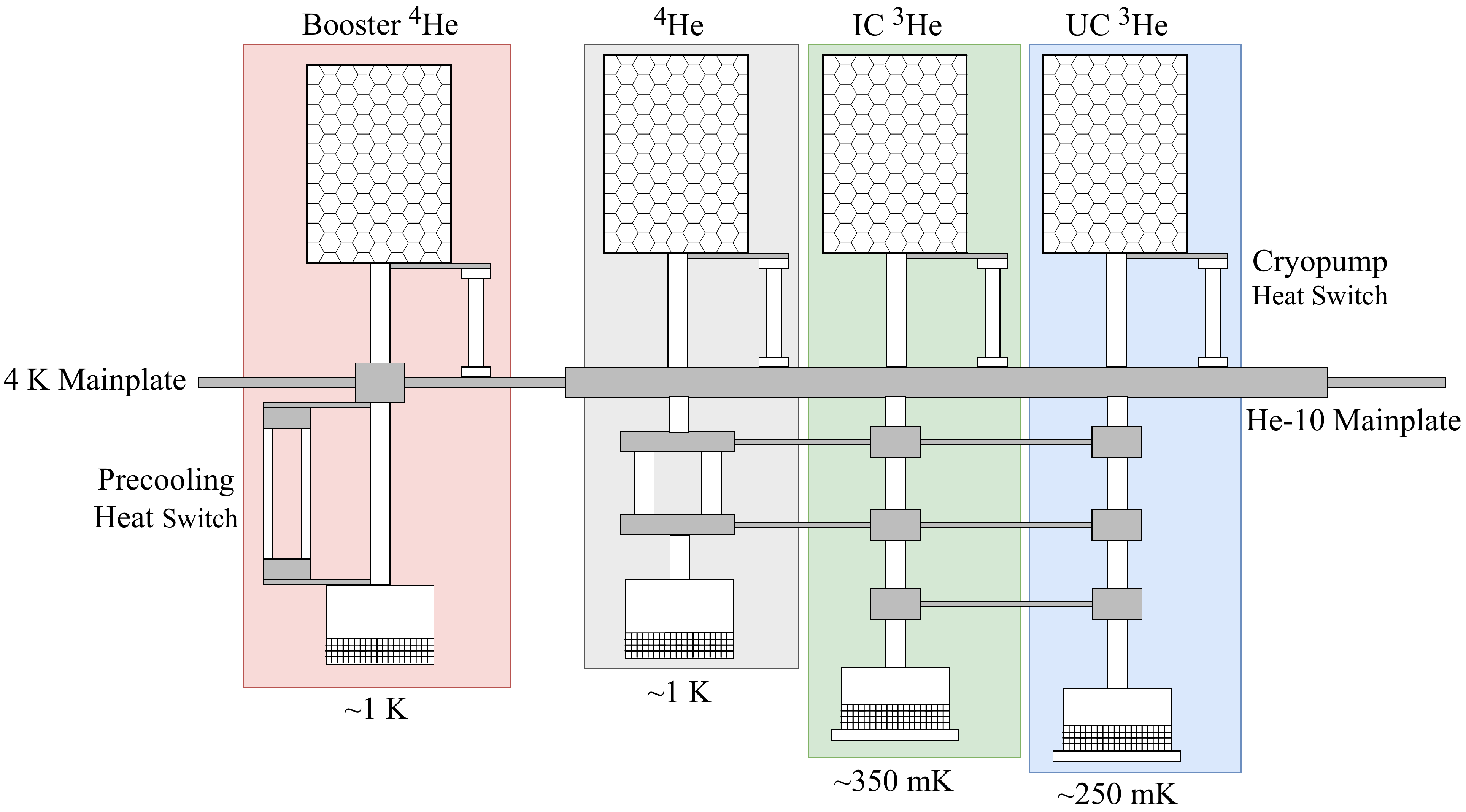}
    \vspace*{2mm}
	\caption{Diagram of the PB-2b millikelvin fridges. The He-10 gas light is comprised of the gray shaded He-4 stage, the green-shaded \emph{Intercold} (IC) He-3 stage, and blue-shaded \emph{Ultracold} (UC) He-3 stage. These stages share a copper mainplate which is in turn bolted to the mainplate of the 4~K shell. The booster He-4 (red shaded stage) is a standalone fridge used to increase the PB-2b hold time. Dark grey shaded components are copper for thermal connections, hexagonal-fill boxes denote the activated charcoal adsorption pumps, and the hatched region on each fridge head denotes the region where liquefied He exists. The He-4 stage of the gas light fridge employs a double-tube architecture which allows convection to more rapidly cool the head and mass attached to it. For the booster He-4 we have employed a similar technology using an external, He-3-charged convective \emph{precooling} heat switch.}
	\label{fig:fridge_diagram}
\end{figure}

\subsection{Millikelvin Refrigeration}
Cooling below 4~K can be achieved using a variety of techniques, including adiabatic demagnetization refrigerators (ADRs), He-3/He-4 dilution refrigerators (DRs), or helium adsorption refrigerators. ADRs are generally avoided in applications requiring SQUIDs due to the strong magnetic fields created by the salt pill. For PB-2b and -2c we have opted to use an helium adsorption refrigerator over a DR due to the fact they are an order of magnitude lower in cost, smaller scale, and easier to integrate, maintain, and operate -- especially at the high altitude, remote site where the Simons Array is located. The PB-2b and -2c adsorption refrigerators are so-called \emph{He-10 gas light} fridges from Chase Research Cryogenics \footnote{\url{http://www.chasecryogenics.com/}}, which are comprised of one He-4, and two He-3 adsorption refrigerator stages ($4+3+3=10$), which reach approximate base temperatures of 1~K, 350~mK, and 250~mK respectively. The enabling physics of all adsorption refrigerators is that of evaporative cooling, which relies on liquefying a quantity of He in the evaporator (or \emph{head}) and then pumping on the liquid to facilitate boil-off and lower the bath temperature.

Adsorption refrigerators not only exhibit load curve behavior in relation to their base temperatures, but are also non-continuous (\emph{single-shot}). When the He reservoir expires (i.e. has completely evaporated) the evaporative cooling ceases, at which point the head temperature of the expired stage increases and the He must be re-liquefied (\emph{recycled}). We define the \emph{hold time} of a stage as the time interval during which the fridge operates below a target temperature without expiring. The configuration of the He-10 gas light, shown in Figure~\ref{fig:fridge_diagram}, is such that the coldest stage --  the UC head -- cannot operate at sufficiently low and constant temperature (i.e. 250~mK) if one of the warmer stages expires first. Thus the practical hold time of a He-10 is determined by the substage with the shortest hold time. For the long exposure observations that Simons Array will conduct \cite{stebor2016simons,arnold2014simons}, it is important to have a hold time which allows for continuous observation for multiple days. The specification for PB-2b and -2c is for the sum of the recycling time $t_{cycle}$ and the hold time $t_{ht}$ to be $t_{tot} = t_{cycle} + t_{ht} \geq72$~hrs (i.e. a three sidereal day schedule). The maximum observation duty cycle from a cryogenic perspective is then $t_{ht} / t_{tot}$ which, for PB-2b and -2c, we are targeting $>$90\%. For reference, PB-1 demonstrated a typical cryogenic duty cycle of $\sim70$\% while operating on a two sidereal day schedule.

Increasing the duty cycle and hold times beyond that of PB-1 is challenging due to the fact the PB-2 receivers are larger with more detectors and wiring, which increases loading to the fridges. Careful consideration and accurate estimation of these loads is important not only in designing the FPT, but also in determining the amount of He required in each stage to reach the desired hold times and duty cycle. In the following sections we describe our work towards characterizing the overall recycling efficiency of the PB-2b and -2c fridges and their hold times which motivates some of the design of the FPT.


\subsubsection{Fridge Cycle Efficiency and Characterization}
Cycling an adsorption refrigerator stage is achieved by first applying electrical power and heating the charcoal cryopump to $\sim$40~K, at which point all He is desorbed. This allows the gas to come in contact with a cold source (the 4~K mainplate) and condense into the evaporator. Initially the gas will be quite hot in comparison to the mainplate and will initially raise the evaporator temperature above that of the mainplate. As the amount of residual gas in the stage being cycled decreases due to condensation, the evaporator temperature can cool to the mainplate temperature. Once the evaporator is sufficiently cold the cryopump is activated: power is turned off and the heat switch linking the cryopump to the mainplate is energized to cool the cryopump, allowing the He bath to evaporatively cool as the boil-off is collected in the cryopump. We define the condensation point $T_{cond}$ as the temperature of the evaporator before cryopumping is initiated. The amount of He that is liquefied is a function of $T_{cond}$, which impacts the stage hold time. We define the cycle efficiency, $\eta$, of a stage as the fraction of the total He charge which remains liquefied in the evaporator once at base temperature, which is just the condensation efficiency while taking into account the self-cooling loss when cooling to base temperature. The cycle efficiency is thus a function of the mass attached to the fridge and can be computed for each configuration. For the purposes of discussing the performance of the He-10 only we neglect this correction.

The two isotopes of He condense at different temperatures: 4.23~K for He-4 and 3.19~K for He-3 \cite{pobell2007matter}, which is sufficiently cold that it is necessary to use an He-4 stage to condense the gas in an He-3 stage so that $\eta_{He-3}$ is near unity. Thus, cycling an He-10 gas light fridge first requires cycling the He-4 stage to condense the He-3 in the IC and UC stage, after which the IC and UC cryopumps are cooled. A typical He-10 gas light cycle is shown in Figure~\ref{fig:frdige_cycle_hold_time_check}.

A model for $\eta$ as a function of cryopump temperature $T_{CP}$ and $T_{cond}$ can be developed  by treating He as an ideal gas in liquid-vapor equilibrium at every point in the stage \cite{cheng1996high}. The total number of moles of He the stage has been filled with is
\begin{equation}
	n = n_l + n_e + n_t + n_{CP}
	\label{eq:total_moles}
\end{equation}
where $n_l$ is the number of moles of liquid, and $n_e$, $n_t$, and $n_{CP}$ the number of gaseous moles in the evaporator, tube and cryopump respectively. Applying the ideal gas law gives
\begin{equation}
	n_l = \frac{n - n_t - \frac{p}{R} \left( \frac{V_{e}}{T_{e}} + \frac{V_{CP}}{T_{CP}} \right)}{1 - (p \tilde{V} / R T_{e})}
	\label{eq:cond_eff}
\end{equation}
with subscripts as in Eq.~(\ref{eq:total_moles}), $p$ the overall system pressure (constant everywhere in the stage), $R$ the ideal gas constant, $V_i$ the volumes corresponding to subscripts $i$, and $\tilde{V}$ the He molar volume. Due to the gradient at location $z$ along the tube between the evaporator and the cryopump, the number of moles in the tube is
\begin{equation}
	n_t = \frac{p}{R} \pi r^2 \int_0^L \frac{dz}{T(z)}
	\label{eq:n_t_of_z}
\end{equation}
where $r$ is the tube radius and $T(z)$ is the temperature in the tube at location $z$. For our fridges, the volume of the tubes is small and we find that $n_t / n$ is less than 0.5\% so we can neglect this in calculating $\eta$. To cool $n$ moles of liquid a temperature $dT$ requires evaporation of $dn$ moles, i.e.
\begin{equation}
	n C(T) dT = L(T) dn,
	\label{eq:cooling_loss}
\end{equation}
where $C(T)$ and $L(T)$ are the specific heat and latent heat of the liquid at temperature $T$. The final number of moles of liquid remaining $n$, after starting with $n_0$ initial moles and cooling the head from temperatures $T_{cond} = T_e$ to $T_{base}$ is
\begin{equation}
	n = n_0 \exp{\int_{T_{e}}^{T_{base}}} \frac{C(T)}{L(T)}dT.
	\label{eq:final_n_selfcooling}
\end{equation}
\begin{figure}[t]
	\centering
	\includegraphics[width = \textwidth]{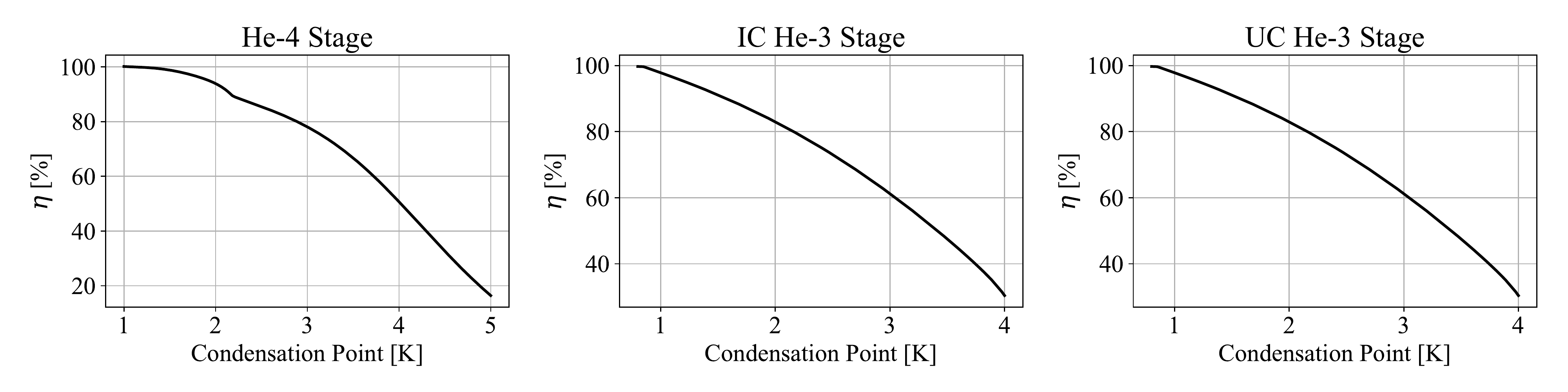}
	\caption{Cycle efficiencies of each stage of the standard He-10 gas light fridge for stage operating base temperatures of 850~mK (He-4), 350~mK (IC He-3), and 250~mK (UC He-3). As the condensation temperature nears the stage base temperature the self-cooling loss becomes negligible and $\eta$ approaches 100\%.}
	\label{fig:cycle_efficiency}
\end{figure}

Although this model is insensitive to pump temperatures over $\sim20$~K, in practice it is necessary to heat the pumps to 45~K before any effect on $\eta$ is negligible. Results of this model are shown in 
Figure~\ref{fig:cycle_efficiency} for the standard He-10 gas light. It is evident that achieving $\eta_{He-3} \sim 100$\% is relatively easy given a typical condensation point of 1.2~K, while a typical $\eta_{He-4}$ will be 70\% in the PB-2b and -2c backends due to the fact that the condensation temperature is bounded below by the  4~K mainplate temperature of 3.2-3.3~K.

\subsubsection{Millikelvin Fridge Characterization}
\label{sec:mK_fridge_testing}
\begin{figure}[t]
	\centering
	\includegraphics[width = \textwidth]{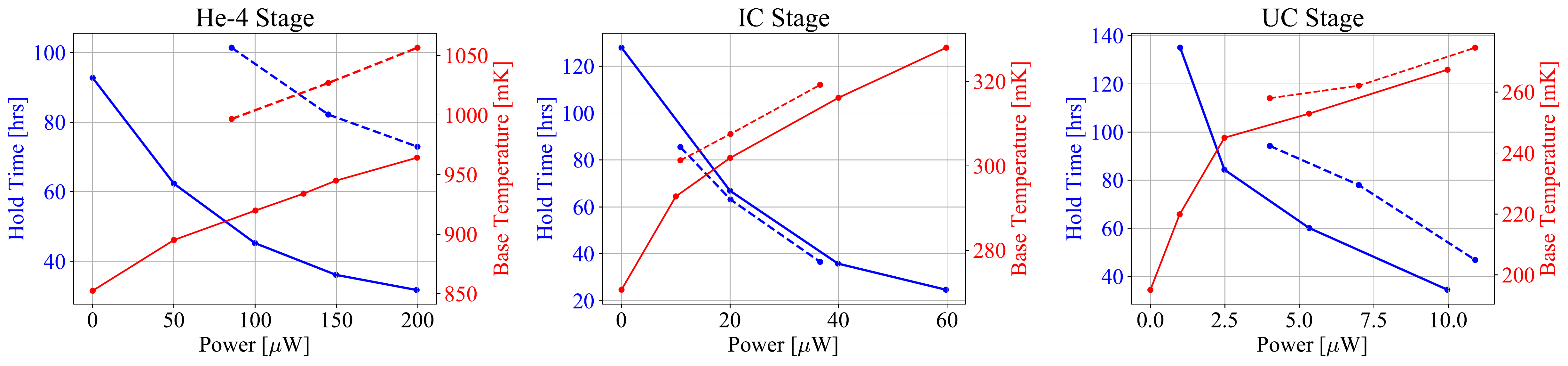}
	\caption{Hold times and base temperatures for the PB-2b (solid lines) and -2c (dashed lines) He-10 fridges. These results indicate that -- for the loading specification of 150~$\mu$W, 20~$\mu$W, and 5~$\mu$W on the He-4, IC, and UC stages respectively -- the expected hold times of PB-2b (-2c) are 36 (81), 63 (61), and 67 (78)~hrs for the He-4, IC, and UC stages respectively. These specifications are over-estimated and we expect to be able to reduce the IC and UC loading sufficiently to allow a 72~hr cycle for PB-2b. The PB-2c gas light is similarly limited by the IC stage to 61~hrs but reduction of the load to this stage quickly allows $t_{ht}$ to exceed 72~hrs.}
	\label{fig:all_gl_plots}
\end{figure}
\begin{figure}[t]
	\centering
    \includegraphics[width = .4\textwidth]{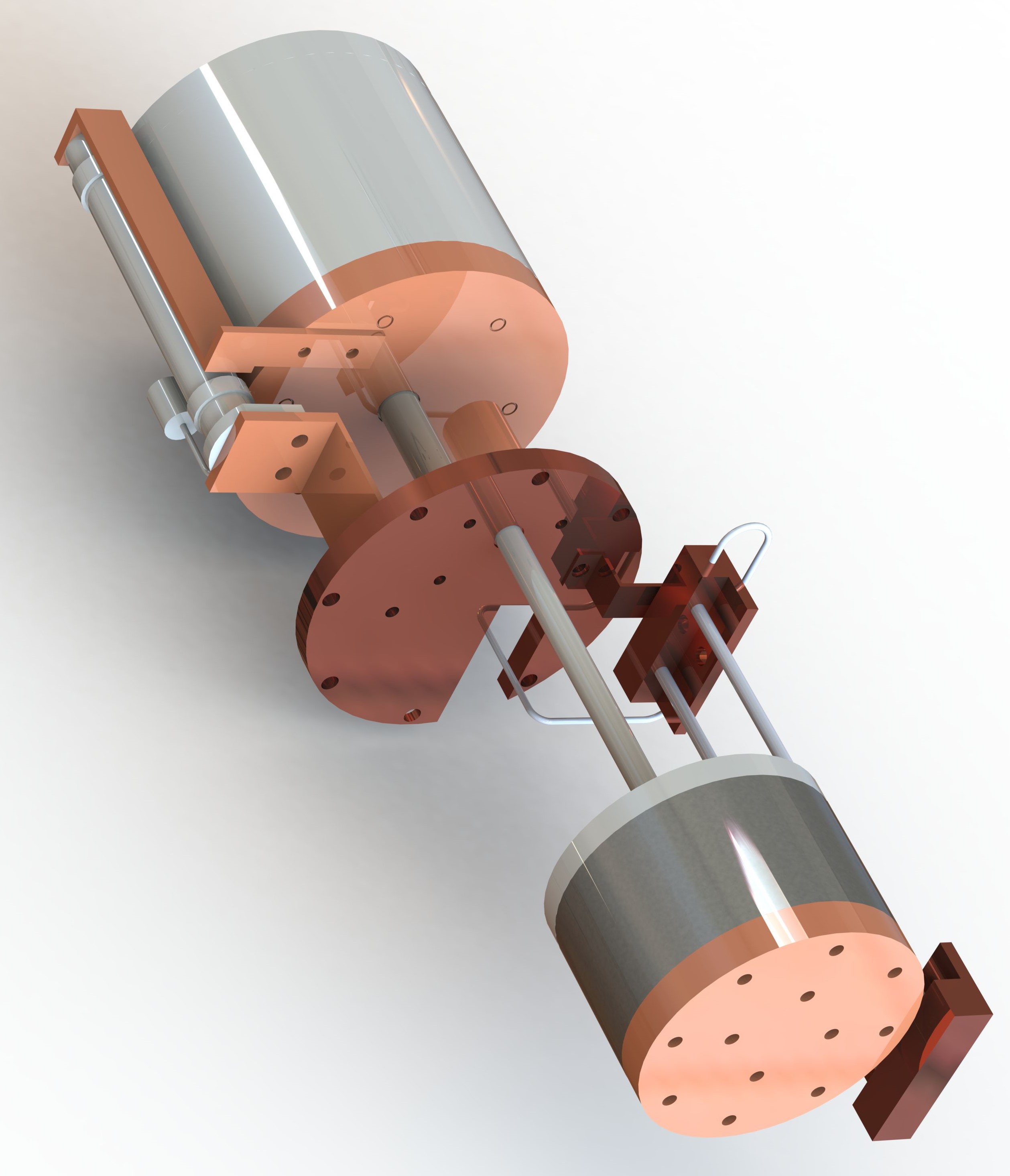}
    \vspace*{2mm}
    \caption{CAD image of the booster He-4 fridge with the cryopump in the upper left corner and the head in the lower right. The precooling heatswitch can be seen to the right of the booster head.}
    \label{fig:booster_cad}
\end{figure}

For PB-2b we chose to purchase the standard model He-10 gas light before millikelvin hardware was finalized so detailed thermal loading estimates did not yet exist. As a rough baseline the specifications for the PB-2b millikelvin loads were defined to be 150~$\mu$W to the 1~K (He-4) stage, 20~$\mu$W to the 350~mK (IC) stage, and 5~$\mu$W to the 250~mK (UC) stage. In order to help inform the design of the FPT we fully characterized this fridge by measuring the load curve and hold times of each stage with no mass attached and no optical loading, as shown in Figure~\ref{fig:all_gl_plots}. With 150~$\mu$W, 20~$\mu$W, and 5~$\mu$W of loading to the He-4, IC, and UC stages the hold times are 36~hrs, 67~hrs, and 63~hrs -- i.e. the practical hold time is 36~hrs and limited by the He-4 stage, which is far below our three sidereal day requirement. Our solution is to install a second standalone He-4 \emph{booster fridge} which meets the hold time requirement. A CAD rendering of the booster fridge is shown in Figure~\ref{fig:booster_cad}. In the new PB-2b configuration the gas light He-4 stage is left unattached to the FPT and is used only as a buffer for the IC and UC stages, which reduces its loading to less than 1~$\mu$W from radiation. Instead, the booster fridge is used to intercept all of the loading incident on the 1~K FPT stage. Owing to the fact that the hold time of the booster fridge is designed to be multiple days, we chose only to verify that it met specifications before integrating in the PB-2b backend so detailed load curve and hold time measurements have not been made.

The booster fridge is charged with 1.3 moles of He-4, has an operating temperature of 936~mK with 150~$\mu$W of loading applied, and demonstrated a hold time of 73~hrs when cycled with a condensation point of 3.7~K in a bare, dark environment. For PB-2c, because the fridge had not already been purchased, we used the He-10 and booster characterization to inform design and fabrication of a custom super gas light (SGL) He-10 from Chase Research Cryogenics. The SGL is an oversized version of the standard gas light He-10 with a larger amount of He-4 and UC He-3 in order to bring the hold time up to 72~hrs in a single fridge. Results from the characterization of the PB-2c SGL are shown in Figure~\ref{fig:all_gl_plots}.

\subsection{Thermal Loading Estimates}
\label{sec:loading_estimates}
\subsubsection{50~K and 4~K Loading}
Modeling thermal loading is important for cryostat design and validation to ensure proper functionality in reaching desired base temperatures. After evacuating the cryostat to $\sim10^{-8}$~Bar there is only conductive loading -- from the mechanical G-10CR shell supports and wiring -- and radiative loading. The heat load $\dot{Q}$ on stage 1 due to stage 2 via conduction of one element is
\begin{equation}
	\dot{Q}_{c} = \frac{A}{l} \int_{T_1}^{T_2} \kappa(T)~dT,
	\label{eq:cond_load}
\end{equation}
where $T_1$ and $T_2$ are the temperatures of the stages, and $A$ is the cross-sectional area, $l$ the length, and $\kappa(T)$ the thermal conductivity of the element connecting the stages. Table~\ref{tab:cond_loads} details the source and contribution of each load to the 50~K and 4~K stages in the PB-2b and -2c backends. Readout and housekeeping wiring from 300~K to 4~K is polyimide-clad, 127$\mu$m diameter manganin alloy. Housekeeping wiring from 4~K to 250~mK is polyimide-clad, 127~$\mu$m diameter niobium-titanium (NbTi) alloy. Readout wiring from 4~K to 250~mK is a polyimide-NbTi superconducting stripline specially developed for the readout system employed in the PB-2 and SPT3G receivers \cite{bender2014multiplexing}, which will be discussed in Section~\ref{sec:mK_loading}. 

\begin{table}[ht]
	\caption{Summary of the 50~K and 4~K conductive loads in the PB-2b and -2c backends. $A$ is the  cross sectional area of one loading component. $N$ is the number of components of each type at each stage. The integrated conductivities are calculated using interpolants of available measured data for each material.}
	\label{tab:cond_loads}
	\begin{center}
	\begin{tabular}{| c | c | c | c | c | c | c |}
	\hline
	\rule[-1ex]{0pt}{3.5ex} Stage & Load Source & $N$ & $A$ [m$^2$] & $ ~ \int \kappa(T)~dT$ [W$\cdot$m] & $l$ [m] & Load [W] \\
	\hline
    \hline
	\rule[-1ex]{0pt}{3.5ex} 50~K & Long G-10 Supports & 6 & 2.68$\times 10^{-5}$ & 97 & 0.22 & 0.07 \\
	\hline
	\rule[-1ex]{0pt}{3.5ex} 50~K & Short G-10 Supports & 24 & 1.18$\times 10^{-5}$ & 97 & 0.047 & 0.59 \\
	\hline
	\rule[-1ex]{0pt}{3.5ex} 50~K & Readout Wiring & 2160 & 1.27$\times 10^{-8}$ & 3950 & 0.057 & 1.9 \\
	\hline
	\rule[-1ex]{0pt}{3.5ex} 50~K & Housekeeping Wiring & 102 & 1.27$\times 10^{-8}$ & 3950 & 0.10 & 0.051 \\
	\hline
	\rule[-1ex]{0pt}{3.5ex} 4~K & Long G-10 Supports & 6 & 2.68$\times 10^{-5}$ & 8.6 & 0.22 & 0.0053 \\
	\hline
	\rule[-1ex]{0pt}{3.5ex} 4~K & Short G-10 Supports & 24 & 1.18$\times 10^{-5}$ & 8.6 & 0.047 & 0.11 \\
	\hline
	\rule[-1ex]{0pt}{3.5ex} 4~K & Readout Wiring & 2160 & 1.27$\times 10^{-8}$ & 258 & 0.049 & 0.14 \\
	\hline
	\rule[-1ex]{0pt}{3.5ex} 4~K & Housekeeping Wiring & 102 & 1.27$\times 10^{-8}$ & 258 & 0.070 & 0.0044 \\
	\hline
	\hline
	\rule[-1ex]{0pt}{3.5ex} \textbf{50~K Total} & & & & & & \textbf{2.6} \\
	\hline
	\rule[-1ex]{0pt}{3.5ex} \textbf{4~K Total} & & & & & & \textbf{0.26} \\
	\hline
	\end{tabular}
	\end{center}
\end{table}

The radiative heat transfer between two bodies at temperatures $T_1$ and $T_2$ with areas $A_1$ and $A_2$, and emissivities $\epsilon_1$ and $\epsilon_2$ is
\begin{equation}
	\dot{Q}_{rad} = \frac{\sigma (T_1^4 - T_2^4)}{(1 - \epsilon_1) / (A_1 \epsilon_1) + 1 /(A_1 F_{1 \rightarrow 2}) + (1 - \epsilon_2) / (A_2 \epsilon_2)}
	\label{eq:rad_load}
\end{equation}
where $F_{1 \rightarrow 2}$ is the \emph{viewing factor} for body 1 and 2 \cite{howell2010thermal}. For two large parallel plates, as in the backend cryostats, $F_{1 \rightarrow 2} = 1$. Using Eq.~(\ref{eq:rad_load}) we can quickly calculate for the PB-2 backend cryostats (with $A_{300~K} = 3.67$~m$^2$, $A_{50~K} = 2.81$~m$^2$) which are made from unpolished aluminum with $\epsilon_1 =  \epsilon_2 = 0.12$, the radiation load to 50~K is in excess of 85~W. This would overwhelm the cooling capacity of the PT415 first stage, which necessitates mitigation of the radiative load. While the calculation for the second stage yields 0.056~W and is within the second stage capacity, it is beneficial to reduce loading as much as possible. This further lowers the 4~K mainplate and assists in the endeavor of maximizing $\eta_{He-4}$.

Mitigation of radiative loading can be achieved by wrapping the 50~K and 4~K shells in multilayer insulation (MLI) blankets which reduces the effective emissivity of the shells. For PB-2b and -2c we chose the Coolcat 2 NW MLI from Ruag \footnote{\url{https://www.ruag.com/en/products-services/space/spacecraft/thermal-systems/cryogenic-thermal-insulation-coolcat}}, which is constructed from many sheets of double-sided, aluminized  (40~nm aluminum thickness on each side) polyester foil with non-woven polyester spacers between adjacent layers (used to reduce conduction between adjacent layers). The full blankets are comprised of multiple stacks of 10-layer blankets which are cut precisely using a laser. The cutting process also bonds each stack by melting the polyester spacers along the laser path. For PB-2b and -2c there are 50 layers at 50~K and 20 layers at 4~K.

MLI reduces radiative loading due to the fact that each internal layer of MLI acts roughly as an isolated radiation shield in radiative equilibrium with its neighbors, with the innermost layer isothermal with the shell it encloses. This would suggest that MLI reduces loading by a factor that is proportional to the inverse of the number of layers. However, this does not take into account conduction between layers or from the interstitial gas, and other non-idealities. Depending on the configuration, i.e. how many edges and seams are required and whether the MLI layers are tightly constrained or allowed to expand, the actual performance of the MLI can be multiple orders of magnitude worse than expected from models built from first principles. To address this, numerous empirical models have been developed to account for these non-idealities \cite{bapat1990performance}. Of these we have chosen the commonly-used Keller model \cite{keller1971thermal}, which gives the heat load per unit area as a sum of the radiative flux
\begin{equation}
	\dot{Q}_{rad} = \frac{A C_r \epsilon}{N_l}(T_h - T_c),
	\label{eq:mli_qdot_rad}
\end{equation}
the contact conduction between layers
\begin{equation}
	\dot{Q}_{c,MLI} = \frac{A C_s \tilde{N}_l^{n_s} T_m}{N_l + 1}(T_h - T_c),
	\label{eq:mli_qdot_cond}
\end{equation}
and conduction from the interstitial gas
\begin{equation}
	\dot{Q}_{gas} = \frac{A C_g p_{int} (T_h^{m + 1} - T_c^{m + 1})}{N_l}.
	\label{eq:mli_qdot_gas}
\end{equation}
$A$ is the area of the cold shell, $N_l$ is the number of layers, $\tilde{N}_l$ is the layer density, $T_m = (T_h - T_c) / 2$, and $p_{int}$ is the interstitial gas pressure. $C_r$, $C_s$, and $n_s$, are parameters dependent on the material and construction of the blankets, and $C_g$ and $m$ are parameters that depend on which gases are present. For our modeling we will consider two contributions of the interstitial mode between 300~K and 50~K: He and N$_2$. The PB-2b and -2c MLI layers have perforations, with a total open area of 0.5-1\%, to help minimize $p_{int}$. In more complicated geometries with many edges and seams, $p_{int}$ can still be multiple orders of magnitude larger than the vacuum vessel pressure and thus $\dot{Q}_{gas}$ can be significant. The interstitial pressures in Table~\ref{tab:mli_params} reflect the PB-2b best-fit values after cryogenic validation from the residual load determined from the PTC capacity map after subtracting $\dot{Q}_{c}$ (Table~\ref{tab:cond_loads}), $\dot{Q}_{rad}$ and $\dot{Q}_{c,MLI}$. We will discuss this further in Section~\ref{sec:50K_4K_vaildation}.

\begin{table}[t]
	\caption{Summary of the values used in the application of the Keller model to the PB-2b and -2c backend cryostats. The units of all parameters are such that the final units of each $\dot{Q}$ are [W]. Values not in parentheses are for He and in parenthesis are for N$_2$.}
	\label{tab:mli_params}
	\begin{center}
		\begin{tabular}{| c | c | c | c | c | c | c | c | c | c |}
			\hline
			\rule[-1ex]{0pt}{3.5ex} & A [m$^2$] & $N_l$ & $C_r \times 10^{10}$ & $C_s \times 10^{8}$ & $\tilde{N}_l$ [cm$^{-1}$] & $n_s$ & $C_g \times 10^{-4}$ & $m$ & $p_{int}$ [mBar] \\
			\hline
			\rule[-1ex]{0pt}{3.5ex} 50~K & 2.81 & 50 & 7.07 & 7.30 & 25 & 2.63 & 4.89~(1.40) & -0.74~(-0.48) & $2.6 \times 10^{-5}$ \\
			\hline
			\rule[-1ex]{0pt}{3.5ex} 4~K & 2.19 & 20 & 7.07 & 7.30 & 20 & 2.63 & 4.89~(1.46) & -0.74~(-0.48) & $1.2 \times 10^{-8}$ \\
			\hline
		\end{tabular}
	\end{center}
\end{table}

\begin{table}[t]
	\caption{MLI contribution to the PB-2b backend loading at 50~K and 4~K.}
	\label{tab:mli_loads}
	\begin{center}
		\begin{tabular}{| c | c | c | c |}
			\hline
			\rule[-1ex]{0pt}{3.5ex}  & $\dot{Q}_{rad}$ [W] & $\dot{Q}_{c,MLI}$ [W] & $\dot{Q}_{gas}$ [W] \\
            \hline
			\rule[-1ex]{0pt}{3.5ex} 50~K & 1.58 & 0.564 & 36.1 \\
            \hline
			\rule[-1ex]{0pt}{3.5ex} 4~K & $7.98 \times 10^{-4}$ & $2.12 \times 10^{-2}$ & $1.88 \times 10^{-2}$ \\
            \hline
        \end{tabular}
    \end{center}
\end{table}

\subsubsection{Millikelvin Loading}
\label{sec:mK_loading}
As with the thermal loads on the PTC thermal intercepts, the loads on the millikelvin refrigerator may be divided into conductive loads and radiative loads.  Three components contribute to the millikelvin conductive loading: the mechanical supports in the FPT, the readout cables, and the RF shield.  By design, thermal radiation from elements internal to the cryostat is negligible, but radiation from the atmosphere that passes through the telescope optical chain is incident on various millikelvin components and contributes non-negligible loading.

\begin{figure}[t]
	\centering
	\includegraphics[width = 0.49\textwidth]{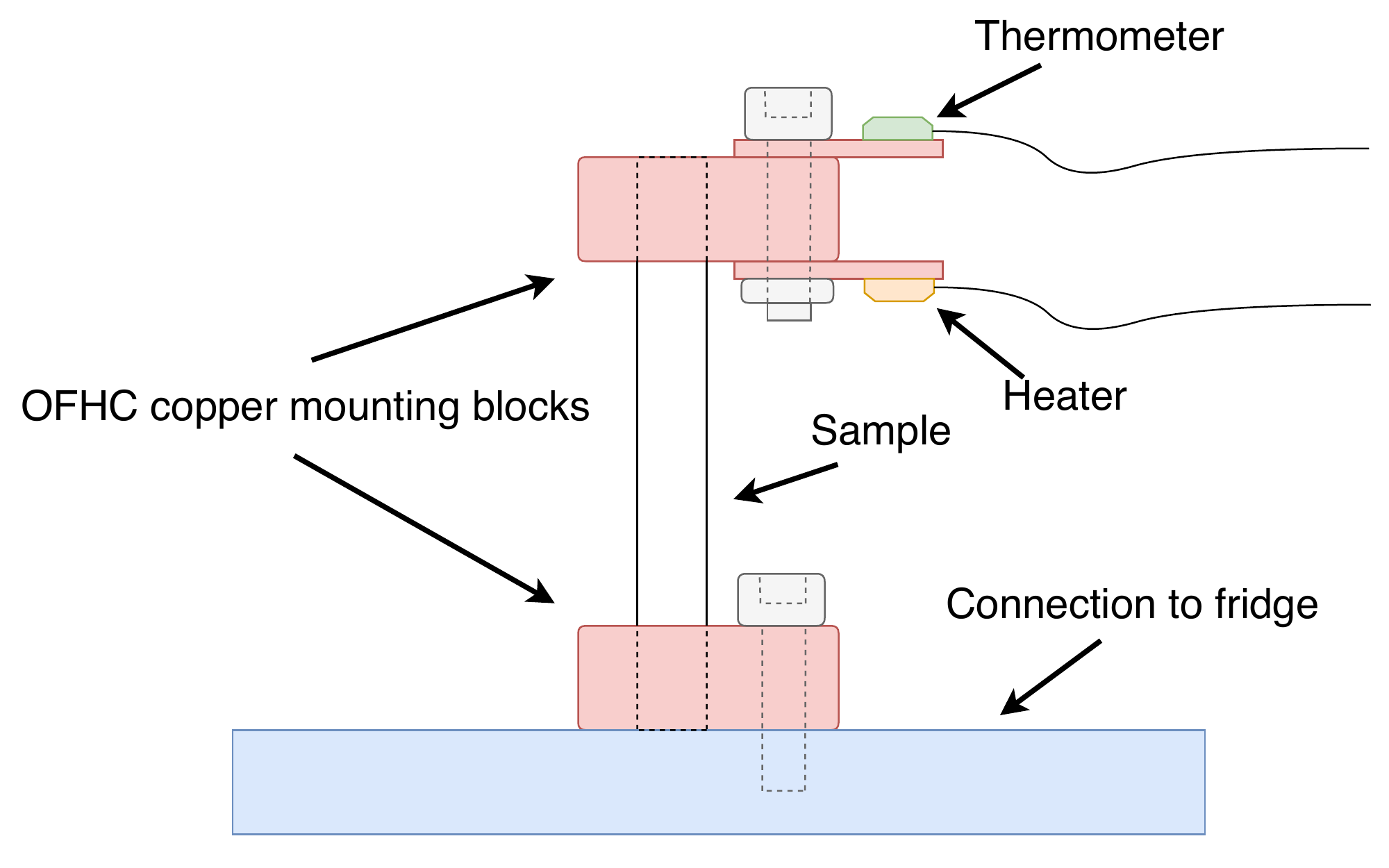}
    \includegraphics[width = 0.49\textwidth]{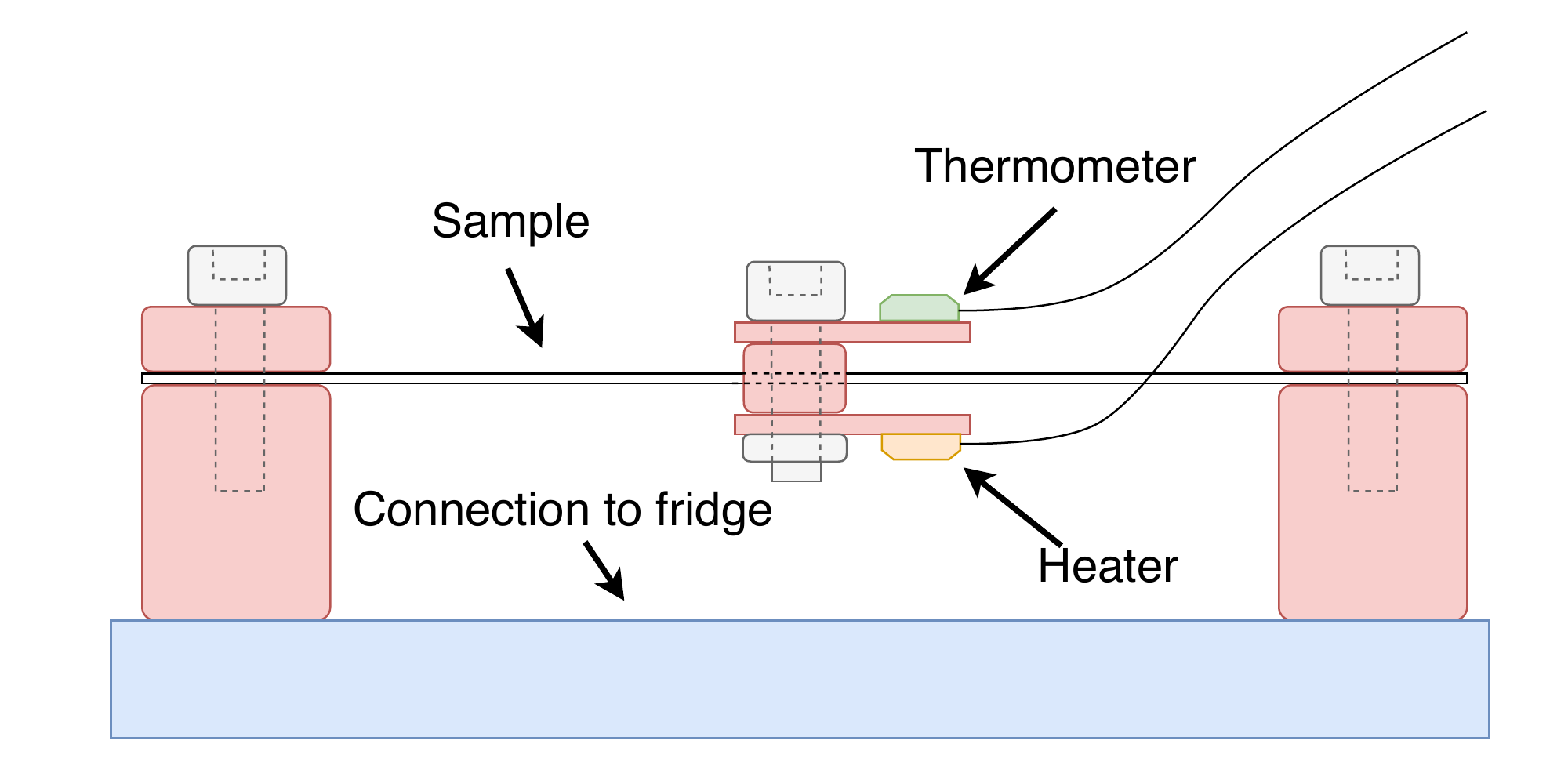}
	\caption{Schematic of the setup for low temperature thermal conductivity measurements with rigid (\emph{left}) and flexible (\emph{right}) materials. When necessary to reduce parasitic thermal loads on the sample samples were also enclosed in a radiation shield.}
	\label{fig:therm_cond_meas_schem}
\end{figure}

The supports of the FPT are a combination of pultruded carbon fiber tubes manufactured by vDijk Pultrusion Products$^*$ \footnote{\url{http://www.dpp-pultrusion.com/}} (DPP) and rods of the commonly used Graphlite pultruded carbon fiber.  The cryogenic thermal conductivity of Graphlite has been well-measured \cite{RunyanJones, Kellaris2014}, but that of DPP was not known at the time of design. To verify this material, measurements of thermal conductivity along the tube axis were performed in the desired temperature range. Known amounts of power $P_{app}$ were applied to one end of a sample and the resulting equilibrium temperature $T_{high}$ was measured, while the other end was fixed at a base temperature $T_{low}$ (see Figure \ref{fig:therm_cond_meas_schem} for a schematic).  The thermal conductivity of DPP was found to be well approximated by a power law
\begin{equation}
	\kappa(T) = \alpha T^\beta,
    \label{eq:mK_cond_fit}
\end{equation}
and the coefficient $\alpha$ and index $\beta$ were obtained by fitting to Eq.~(\ref{eq:cond_load}).  Due to cooling power limitations of the adsorption fridge used for testing, measurements were performed with $T_{low} \sim 300$~mK from a single-shot He-3 adsorption refrigerator, and separately with $T_{low} \sim 1.2$~K from pumped liquid He-4.  The best fit in the 0.25~K---2~K range is $\kappa_{DPP}(T) = 4.17~T^{1.21}$~mW/m$\cdot$K, and the best fit in the 1.4~K---4.5~K range is $\kappa_{DPP}(T)~=~7.59~T^{0.61}~$mW/m$\cdot$K.

The readout cables are fabricated from a custom stack-up of polyimide, superconducting NbTi, photoresist, and adhesive layers. Although cryogenic thermal conductivities of polyimides and NbTi have been measured \cite{Barucci2000, Olson1993}, building an accurate thermal model of the combined cable is subject to large errors due to uncertainty in the properties of the remaining materials.  Thus, a measurement of the cryogenic thermal conductivity of the readout cables was performed in a manner similar to that described above.  Since the cables are flexible, they were held taut and clamped on opposite sides at $T_{low}$, while a central clamp with a heater and thermometer was used to apply power (see Figure~\ref{fig:therm_cond_meas_schem}).  As the cross-sectional area of the cables is fixed but the distance between thermal intercepts in the PB-2 backends is not, the relevant quantity of interest is the thermal conductance per unit length, $G/L$.  The best fit in the 0.25~K---1.5~K range is $G/L = 7.1~T^{1.79}$~$\mu$W/mm$\cdot$K, and the best fit in the 1.4~K---6~K range is $G/L = 9.3~T^{1.05}$~$\mu$W$\cdot$mm/K.

The RF shield is comprised of 300~$\text{\normalfont\AA}$ of aluminum deposited on a 6.35~$\mu$m sheet of polyethylene terephthalate.  The strong dependence of the cryogenic thermal conductivity of aluminum films on their purity motivated a measurement of a sample of the RF shield in a manner similar to that of the readout cables.  Due to the sample's long time constant for equilibration and fridge hold time limitations, measurements were only performed in the 1.4~K---6~K range.  Since the thermal conductivity will decrease more strongly below the critical temperature of the aluminum film, extrapolating measurements from this range provides an overly pessimistic estimate of the thermal loads at colder temperatures.  Moreover, as the RF shield extends radially as well as vertically away from the focal plane, the relevant quantity of interest is the thermal conductance multiplied by the thickness $d$ of the shield, $\kappa d$.  The measured best fit is $\kappa d = 33.5~T^{1.41}~$nW/K.

Optical loads which contribute to the thermal loading of the adsorption fridge include out-of-band radiation absorbed by the 350 mK metal mesh filter and in-band radiation absorbed by emissive lenslets on the focal plane. The particular estimation of these loads depends on the specifics of the optical design (and therefore differ for PB-2b and -2c), which will be detailed in an upcoming publication. For the purposes of estimating the thermal budget, conservative upper limits are quoted here.

A summary of the expected thermal loads on the He-10 fridge substages is given in Table \ref{tab:fridge_loads}.  All conductive loads were calculated according to Equation~(\ref{eq:cond_load}) with known geometries and with temperatures measured from a cooldown with realistic thermal loads applied to each fridge intercept via resistive heaters.

\begin{table}[t]
	\caption{Projected thermal loads [$\mu W$] on the fridge intercepts for PB-2b~(PB-2c)}
	\label{tab:fridge_loads}
	\begin{center}
		\begin{tabular}{| c | c | c | c | c | c |}
			\hline
			\rule[-1ex]{0pt}{3.5ex} & Mechanical supports & Readout cables & RF shield & Radiation & Total\\
            \hline
            \rule[-1ex]{0pt}{3.5ex} He-4 stage & 13.8 & 40.6 & 21.3 & 0~(0) & 76.1~(76.1)\\
            \hline\rule[-1ex]{0pt}{3.5ex} IC stage & 0.8 & 1.8 & 0.9 & $<$4.8~($<$6.7) & $<$8.3~($<$10.2)\\
            \hline
            \rule[-1ex]{0pt}{3.5ex} UC stage & 0.3 & 0.1 & 0.1 & $<$1.7~($<$3.5) & $<$2.0~($<$3.8)\\
            \hline
        \end{tabular}
    \end{center}
\end{table}
\section{Commissioning the PB-2b and -2c Backends}
We chose to construct and validate the PB-2b and -2c backends in series with essentially no changes in design or construction, excepting the SGL He-10 fridge specifically designed to have increased hold times beyond that of the PB-2b standard gas light He-10. Apart from this, cryogenic performance for the two backends is similar so the majority of the data shown in the following sections is only for the PB-2b backend.

\subsection{50~K and 4~K Shell Construction}
The PB-2b and -2c 50~K and 4~K shells are constructed from aluminum 1100 (Al1100) alloy panels which are 3 mm and 4 mm thick respectively. These panels are bolted to an Al6061 alloy frame with Apiezon N grease between the panels and the frame at 4~K (no interface material exists at 50~K). This is an acceptable construction at 4~K in the limit that the only load that needs to be transferred to the PTC is the relatively small radiative load. In order to both minimize thermal gradients across the panels and preserve a lightweight construction, the conductance of the panels was augmented with very high purity metal ribbons.

Aluminum is a good choice for both 50~K and 4~K temperature ranges due to the fact that its thermal conductivity can be $\sim10^4$~W/m$\cdot$K or higher \cite{woodcraft2005recommended}. 75~mm~$\times$~0.5~mm 99.9998\% purity (6N) aluminum ribbons were annealed in an N$_2$ environment at 300~$^o$C for 8~hours and were attached to the panels using Stycast 2850ft epoxy. For the front, rear, and bottom panels we placed ribbons only along the edges on one side. For the top, mainplate, and trapezoidal panel (seen in Figure~\ref{fig:backend_exploded} between the mainplate and top panels) we chose to cover the entire surface of both sides in 6N ribbons. Initial stress testing via rapid thermal cycling from 300~K to 77~K revealed that surface preparation is critical to ensure a robust, high quality thermal connection between the Al1100 panel and the 6N aluminum. In order to create the best quality interface, both the panels and ribbons were first roughened using 60-grit sandpaper, then a toluene-based epoxy adhesion-promoter \footnote{\url{https://www.lord.com/products-and-solutions/chemlok-ap-134-primer}} was applied, after which the epoxy (prepared with catalyst 9 \footnote{\url{http://na.henkel-adhesives.com/product-search-1554.htm?nodeid=8797863247873}}) was applied to the panel, and then the 6N ribbon.

We chose to use an initially thick layer of epoxy ($\sim$2~mm) and apply an even pressure of $\sim10$~kPa along the ribbon to ensure complete coverage and adhesion while keeping the epoxy layer thin by forcing out the excess. These epoxy interfaces were cured at room temperature for 24~hrs with the 10~kPa pressure applied, after which the final epoxy thickness is consistently 0.5~mm. Attaching 6N aluminum ribbons lowers thermal gradients across the shell panels to their practical minima for the geometries in the PB-2b and -2c backends. The remaining gradients between the coldheads and shell sections are subsequently dominated by the thermal interface resistance between the panels, and the Al6061 frame and its lower conductivity. However, the small radiation and G-10CR conduction heat loads, and large parallel heat path at these interfaces means these gradients are typically about 0.5~K or less.

For the large conductive load from the SQUID wiring harnesses it is necessary to implement a more carefully engineered solution with special consideration given to the interfaces between perpendicular panels. This is especially important because the performance of the SQUID pre-amplifiers is highly temperature dependent. Avoiding local hotspots due to the wiring harness load and creating the coldest stage possible minimizes complications in readout of the focal plane and increases mapping speed. In order to more effectively transmit the wiring harness load across the two shell interfaces between the top panels and the shell mainplates, we completely covered these panels on both sides in the 6N aluminum ribbons and have added extra length. These lengths are then clamped to the adjacent panel using stainless steel M5 screws, washer/nut plates, and split ring lock washers. Screws are tightened to $\sim80$\% their yield stress. At the interface between the 4~K mainplate and the trapezoid panel there is not enough space to accommodate this solution so we have implemented a C101 L-bracket instead of the clamped overhanging 6N ribbons.

\subsection{Heatstraps}
\begin{figure}[t]
	\centering
	\begin{subfigure}[h]{.43\textwidth}
		\includegraphics[width=\textwidth]{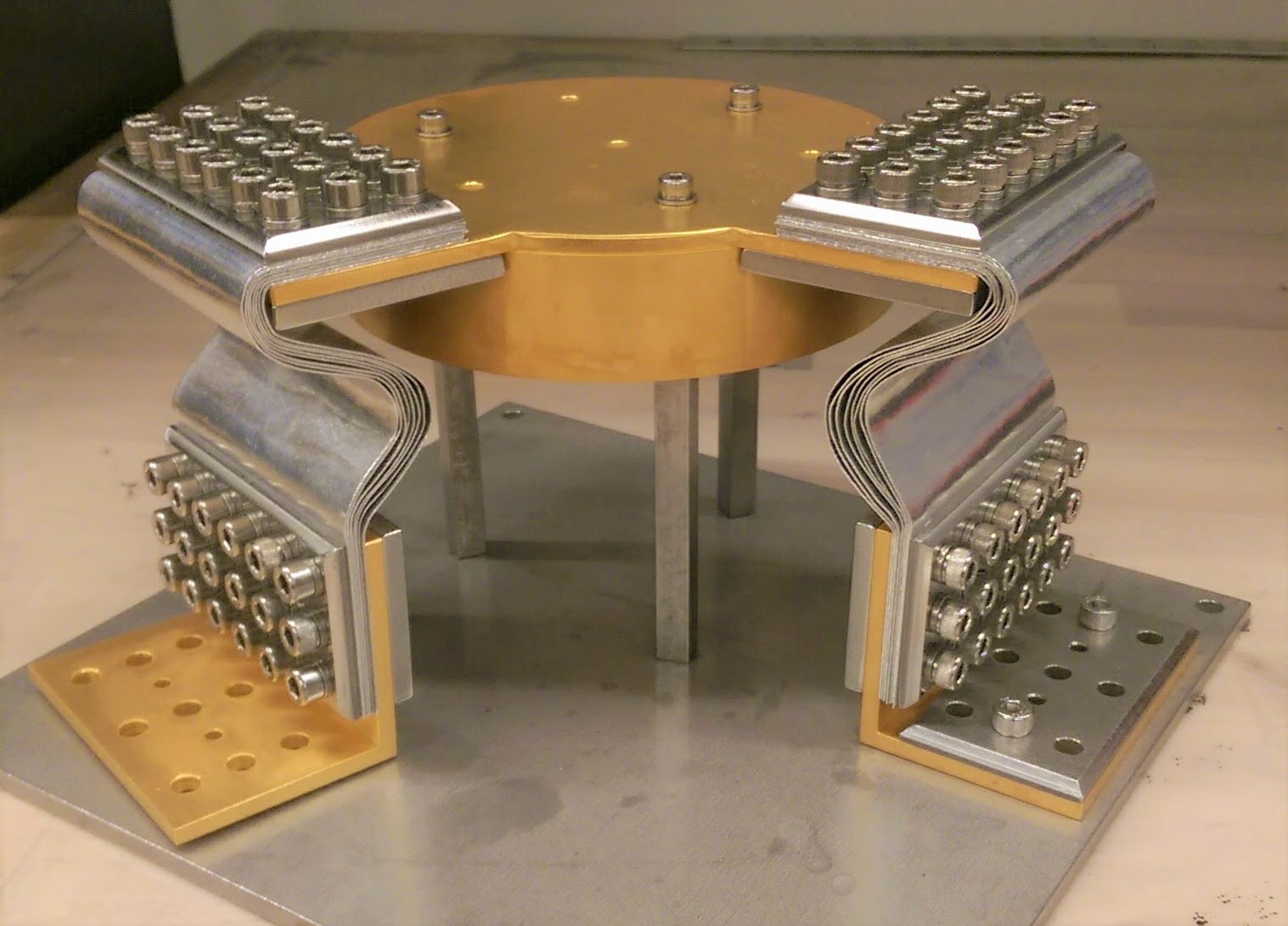}
		\label{fig:4K_mp_ch_heatstrap}
	\end{subfigure}
	\begin{subfigure}[h]{.545\textwidth}
		\includegraphics[width=\textwidth]{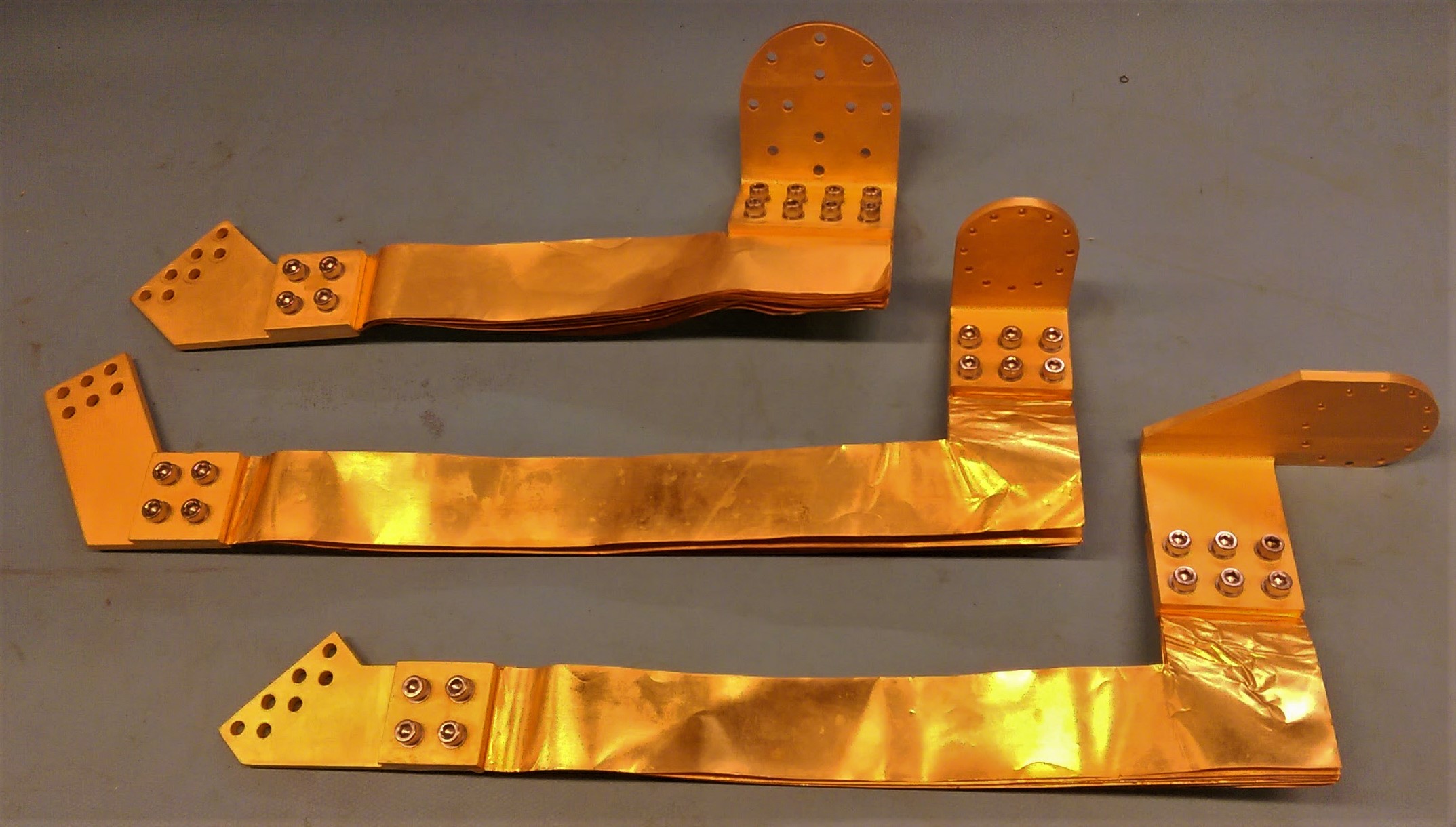}
		\label{fig:mK)_heatstraps}
	\end{subfigure}
	\caption{(\emph{Left}) The PB-2b mainplate-coldhead heatstrap consists of two sets of eight 75~mm~$\times$~160~mm 6N aluminum ribbons which are clamped to gold-plated C101 adapters on each end. Each clamped interface uses 24 M5 stainless steel screws and split-ring lock washers with stainless steel pressure-spreading \emph{washer plates} and \emph{nut plates}, and no thermal interface material. (\emph{Right}) The PB-2b millikelvin heatstraps after gold-plating. Pressure is applied to the FPT stages using 10-32 screws and the same washer and nut plate method as the 4~K mainplate-coldhead heatstrap, and to the millikelvin fridge heads using brass 4-40 screws with serrated Belleville washers.}
	\label{fig:heatstraps}
\end{figure}

One of the most important components of any large cryostat is the heatstrap connecting the PTC to the mainplates because it must pass the entire stage's loading. It is rarely possible to directly mount critical components to the coldheads so the lowest temperature presented to the He-10 fridge and SQUIDs is correspondingly above the PTC base temperature. For the PB-2b and -2c backends the 50~K mainplate-coldhead heatstrap is less critical because the main function of the 50~K shell is as a radiation buffer and an adequate base temperature of the 4~K shell is relatively insensitive to a mean 50~K shell temperature drift of $\pm$10~K. The 50~K mainplate-coldhead heatstrap is simply constructed of six stacks of eight 20~mm~$\times$~80~mm pieces of 6N aluminum ribbons which are clamped to the PTC coldhead and 50~K mainplate. This implementation does not optimize the surface area or interfaces of the thermal connection. The 4~K mainplate-coldhead heatstrap (Figure~\ref{fig:heatstraps}) is more carefully designed in order to minimize the thermal contact resistance between the coldhead and mainplate. While using 6N aluminum is advantageous due to it high bulk conductivity at 4~K (higher than that of C101 when the 6N is annealed), there is significant concern in accessing this due to the robust aluminum oxide which rapidly forms at room temperature in atmospheric conditions \cite{krueger1972initial}. This problem is exacerbated when multiple 6N-6N interfaces are present. Our solution is to prepare the joints with very high bolt force in order to fracture the oxide and to deform the soft 6N aluminum (enhances the effective cross sectional area of the joint). We do not disassemble these interfaces so as to avoid re-oxidation. Rather, the interface which is dissembled is the gold-plated copper interface to the coldhead and mainplate.

Due to the fact that the thermal interface (Kapitza) resistance \cite{pollack1969kapitza} scales as $T^{-3}$ we have paid special attention to optimizing the conductance of the millikelvin heatstraps by minimizing the number of bolted interfaces \cite{berman1956some, salerno2003thermal}. Of additional concern is the potential for vibrations coupling to millikelvin stages which can cause microphonic heating and is deleterious to the millikelvin base temperatures. The PB-2b and -2c millikelvin heatstraps are constructed from C101 \emph{feet}, that bolt to the fridge heads and FPT stages and have 15 layers of 0.1~mm thick C101 copper ribbons bolted between C101 pressure plates and the feet. These ribbons are then welded along the edges so there is a bulk thermal connection to the feet in addition to the contact interface of each ribbon. A picture of the PB-2b millikelvin heatstraps is shown in Figure~\ref{fig:heatstraps}. As will be mentioned in Section~\ref{sec:mK_commissioning}, we have reason to believe the conductance of these heatstraps is limited by the thermal interface resistance between the feet and the FPT and He-10 stages.

\subsection{50~K and 4~K Cryogenic Validation}
\label{sec:50K_4K_vaildation}
To first-order, cryogenic validation of the PB-2b and -2c backend 50~K and 4~K stages simply entails ensuring that an acceptable base temperature is reached; i.e. $\sim$50~K and $\sim$3.5~K. However, encompassed in this is a careful accounting of the observed thermal loads and a comparison with the model to ensure significant spurious loads are not observed. Additionally we require that thermal time constants are low enough that egregiously long cooldown times are not observed, but this is generally solved in tandem with the optimization of thermal gradients as $\tau_{thermal} \sim G^{-1/2}$ where $G$ is the thermal conductance linking a body to the cold source. 

\begin{figure}[t!]
	\centering
	\includegraphics[width = .8\textwidth]{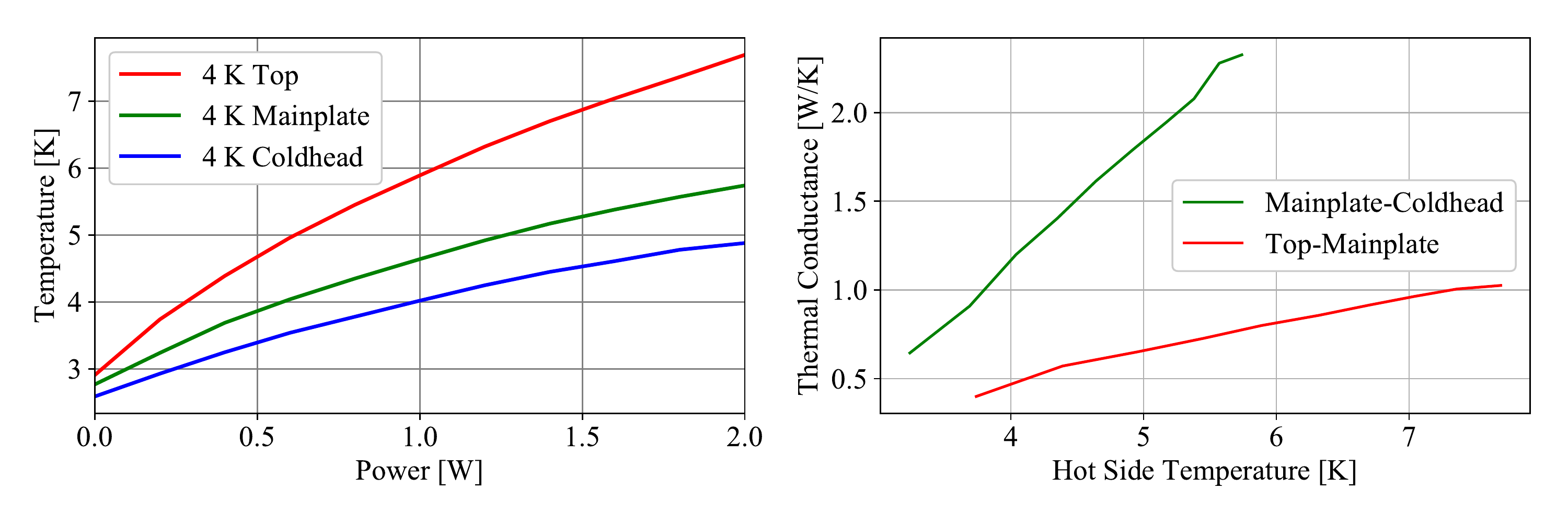}
	\caption{(\emph{Left}) Load curve taken in the first run of the PB-2b backend using a heater on the top panel. (\emph{Right}) Measured thermal conductance of the mainplate-coldhead and the top-mainplate heatstrap.}
	\label{fig:run01b_load_curve}
\end{figure}

\begin{figure}[t!]
	\centering
	\includegraphics[width = .8\textwidth]{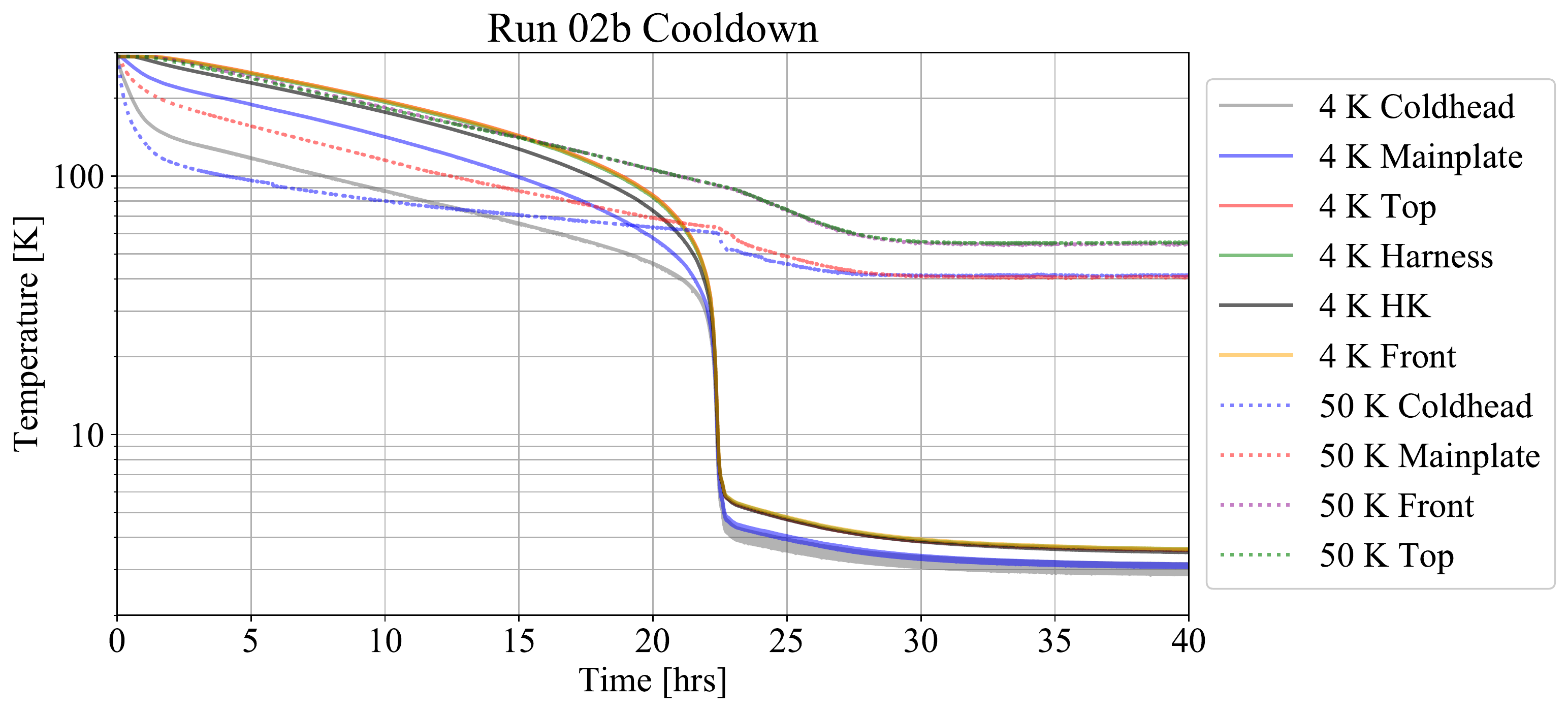}
	\caption{Cooldown plot of the second cooldown of the PB-2b backend. Base temperature is reached in $\sim$35~hrs.}
	\label{fig:run02b_cooldown}
\end{figure}

\begin{figure}[t!]
	\centering
	\includegraphics[width = \textwidth]{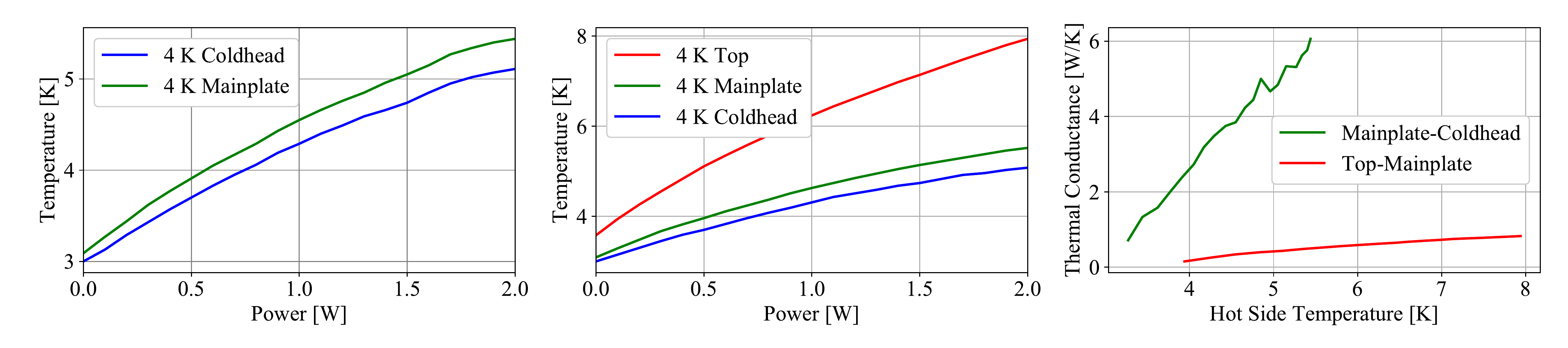}
	\caption{(\emph{Left}) Run 02b load curve taken using a heater on the 4~K mainplate. This is a more accurate measure of the mainplate-coldhead heatstrap $G$. (\emph{Center}) Run 02b load curve taken using a heater on the 4~K top panel. (\emph{Right}) Heatstrap thermal conductances as a function of the hot side temperature.}
	\label{fig:run02b_load_curve}
\end{figure}

For the PB-2b backend, cryogenic validation was done in two cooldowns. The first cooldown (run01b) was started on August 23rd, 2016, without any of the three wiring harnesses in order to measure the 4~K load curve, 50~K base temperature, and evaluate the performance of all heatstraps and thermal interfaces. Measurement of the 4~K load curve and 50~K base temperature with no harnesses installed allows us to precisely measure the loading due to the harnesses after installation and evaluate whether their thermal performance is acceptable. Figure~\ref{fig:run01b_load_curve} shows the PTC second stage load curve taken using a heater on the top panel of the 4~K shell and the resultant thermal conductance $G$ for the thermal connection between the mainplate and the coldhead, and between the top panel and the mainplate.

Figure~\ref{fig:run02b_cooldown} shows the second cooldown (run~02b), which began October 28th, 2016, of the PB-2b backend after installing the three wiring harnesses and re-mating the mainplate-coldhead and top-mainplate heatstraps at higher screw torques and using Apiezon~N grease as a thermal interface material. Load curves and the corresponding $G$ for the two heatstraps in run~02b are shown in Figure~\ref{fig:run02b_load_curve}. From this we see that the both the mainplate-coldhead $G$ and the top-mainplate $G$ have been improved, although the increase is more significant in the former. We estimate that recycling the He-10 fridge adds $\sim$150~mW of loading ($\Delta T_{MP} \sim 100$~mK) to the 4~K mainplate so all efforts in minimizing the mainplate-coldhead gradient will aid in maximizing $\eta$. The PB-2c 50~K and 4~K validation was also achieved on the second run (run~02c). Base temperatures for both 50~K and 4~K validation runs on PB-2b and -2c are shown in Table~\ref{tab:base_temps}.

\begin{table}[t]
	\caption{Base temperatures [K] for the PB-2b and -2c backends during their 50~K and 4~K validation run.}
	\label{tab:base_temps}
	\begin{center}
		\begin{tabular}{| c | c | c |}
			\hline
			\rule[-1ex]{0pt}{3.5ex} Location & PB-2b & PB-2c \\
			\hline
			\rule[-1ex]{0pt}{3.5ex} 4 K Coldhead & 3.00 & 2.81 \\
			\hline
			\rule[-1ex]{0pt}{3.5ex} 4 K Mainplate & 3.09 & 3.05  \\
			\hline
			\rule[-1ex]{0pt}{3.5ex} 4 K Top & 3.55 & 3.47 \\
			\hline
			\rule[-1ex]{0pt}{3.5ex} 4 K Front & 3.57 & 3.55 \\
			\hline
			\rule[-1ex]{0pt}{3.5ex} 4 K Bottom & -- & 3.47 \\
			\hline
			\rule[-1ex]{0pt}{3.5ex} 4 K Harness & 3.59 & -- \\
			\hline
			\rule[-1ex]{0pt}{3.5ex} 50 K Coldhead & 40.9 & 32.6 \\
			\hline
			\rule[-1ex]{0pt}{3.5ex} 50 K Mainplate & 41.1 & 39.3 \\
			\hline
			\rule[-1ex]{0pt}{3.5ex} 50 K Top & 56.1 & 50.1 \\
			\hline
			\rule[-1ex]{0pt}{3.5ex} 50 K Front & 55.4 & 50.4 \\
			\hline
		\end{tabular}
	\end{center}
\end{table}
Base temperatures for run~02b and run~02c are listed in Table~\ref{tab:base_temps}. From these values and the PTC capacity map \cite{green2015cooling} and load curve in Figure~\ref{fig:bare_pb2b_pt415_load_curve} it is evident that the 50~K loading for PB-2b (-2c) is $\sim$40~W ($\sim$20~W) and the 4~K loading is $\sim$0.30~W ($\sim$0.25~W). Our modeling of the conductive loads in Section~\ref{sec:loading_estimates} is consistent with these values and the discrepancy between the two backends can be attributed to more thorough pumping out of the PB-2c backend than PB-2b for their respective runs. If we attribute the remainder of the observed loads to $\dot{Q}_{gas}$ and assume the pressures for He and N$_2$ are equal, we can obtain rough estimates for the interstitial gas pressure during the cooldown of each cryostat. For run~02b (run~02c) we have at 50~K $p_{int} \sim 2.6 \times 10^{-5}~(\sim 1.1 \times 10^{-6})$~mBar and at 4~K $p_{int} \sim 1.2 \times 10^{-8}~(\sim5.0 \times 10^{-9})$~mBar. For reference the vacuum vessel pressure gauge typically reads $10^{-5}-10^{-6}$~mBar so these numbers are consistent at 50~K. The interstitial pressures for 4~K are unphysically low, most likely due to uncertainties in the thermal loading model of other components. Regardless, no large spurious loads or gradients were observed during run~02b and run~02c. This, in conjunction with the verification that the 4~K mainplate in both cryostats will reach a low enough base temperature such that $\eta_{He-4} \sim$70\% completes the 50~K and 4~K validation of the PB-2b and -2c backends.

\subsection{Millikelvin Validation}
\label{sec:mK_commissioning}
\begin{figure}[t]
	\centering
	\includegraphics[width = .8\textwidth]{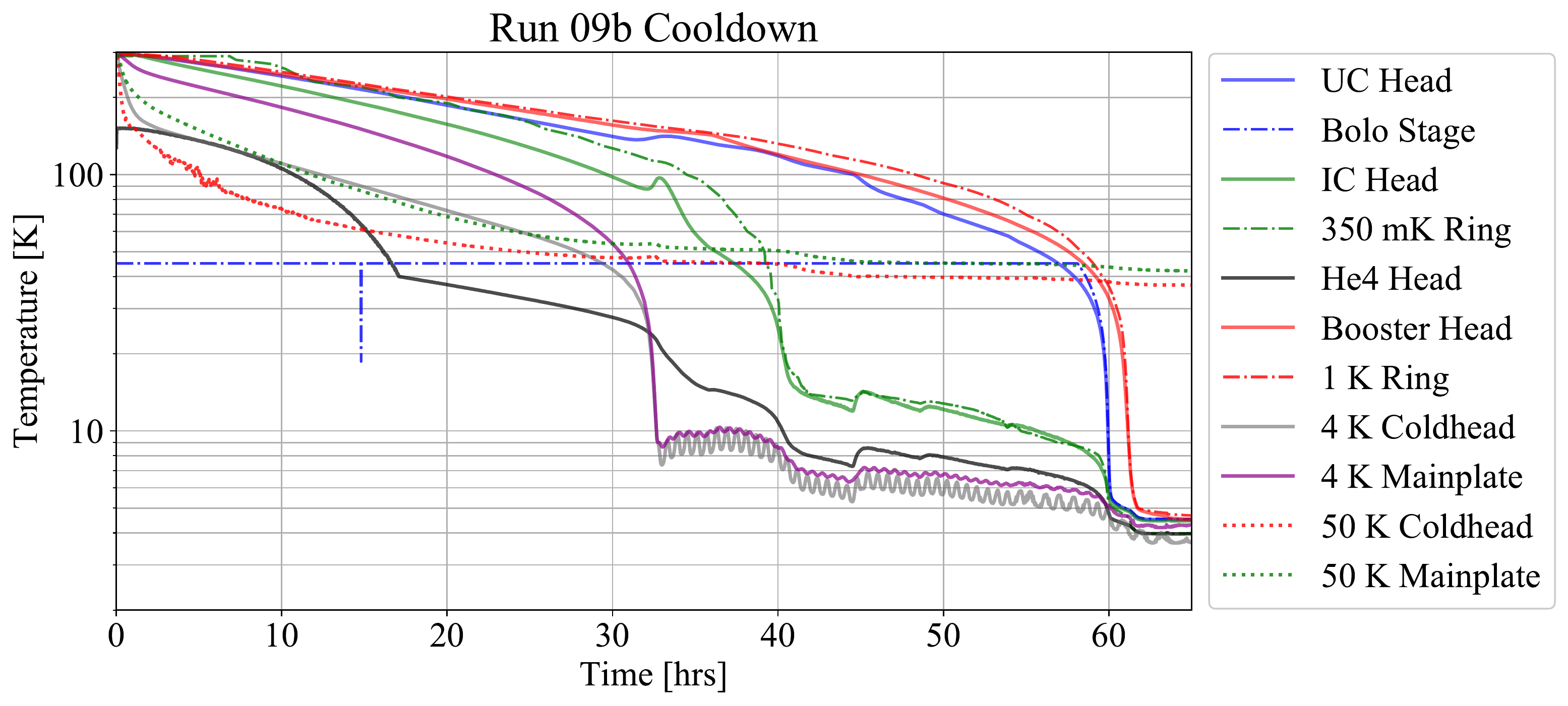}
	\caption{Cooldown plot of run 09b of the PB-2b backend with the bare FPT installed (no detector modules). Base temperature is reached in $\sim$65~hrs.}
	\label{fig:run09b_coldown}
\end{figure}
\begin{figure}[t]
	\centering
	\includegraphics[width = \textwidth]{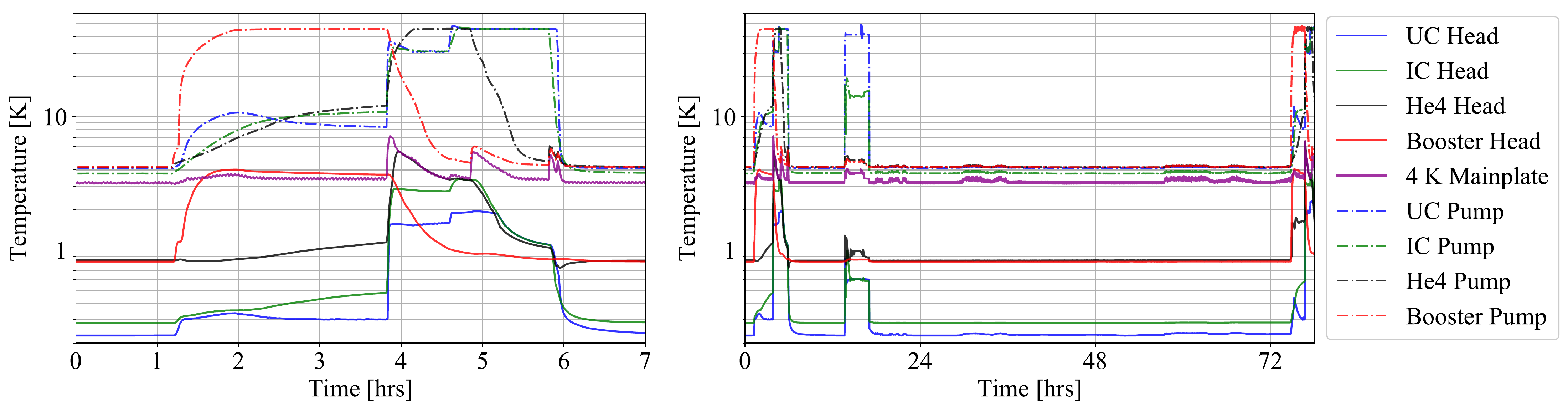}
	\caption{(\emph{Left}) Plot showing PB-2b He-10 fridge recycling which takes 6~hrs to reach a condensation point low enough that $t_{tot}$ is in excess of 72~hrs. The booster fridge is cycled before the He-10 in order to keep the 4~K mainplate temperature as cold as possible so $\eta_{He-4}$ is maximized. (\emph{Right}) Fridge cycle and hold time plot during initial focal plane integration demonstrating that $t_{tot}$ of the PB-2b fridge is $\geq 72$~hrs with $\sim$75, 0, 6, and 1.5~$\mu$W of loading to the booster, He-4, IC, and UC stages. The portion of the plot between 14 and 20~hrs is deliberate heating of the stage which is performed during various detector and readout commissioning tests. The fridge cycle was started before expiration of any fridge stages so the hold time is actually longer than shown here.}
	\label{fig:frdige_cycle_hold_time_check}
\end{figure}

\begin{table}
	\caption{Summary of the values used to obtain the millikelvin heatstrap $G$. $\Delta T$ refers to the temperature gradient from the fridge head to the corresponding FPT stage. In this way $G$ is a measurement of the combination of the thermal contact interfaces on both ends of each heatstrap, as well as the bulk properties.}
	\label{tab:base_temps}
	\begin{center}
		\begin{tabular}{| c | c | c | c |}
			\hline
			\rule[-1ex]{0pt}{3.5ex} Heatstrap & $\Delta T$ [mK] & Power [$\mu$W] & $G$ [$\mu$W/mK]  \\
			\hline
			\rule[-1ex]{0pt}{3.5ex} Booster & 52 & 68 & 1.3 \\
			\hline
			\rule[-1ex]{0pt}{3.5ex} IC & 21 & 18 & 0.85 \\
			\hline
			\rule[-1ex]{0pt}{3.5ex} UC & 3 & 3.8 & 1.26 \\
			\hline
		\end{tabular}
	\end{center}
\end{table}

Once the 50~K and 4~K validation is complete we begin integration of the millikelvin fridges and FPT. It is advantageous to integrate the bare (no attached mass) millikelvin refrigerators in order to identify any stray loading (such as light leaks) that may be present, and to diagnose any deleterious thermal gradients between the 4~K mainplate and the fridge condensers. This required multiple cooldowns for the PB-2b backend due to large gradients between the booster condenser and the 4~K mainplate, and long recycling times due to improper thermal anchoring of the precooling switch. Figure~\ref{fig:run09b_coldown} shows the cooldown for run~09b (started on June 9th, 2017) in which the millikelvin validation was completed. This run contained only the mechanical structure of the FPT, i.e. no detector modules, MMFs, or RF shields were installed. For the PB-2b and -2c backends millikelvin validation entails demonstrating acceptable base temperatures (and identifying spurious loads), fridge hold time, and fridge recycling time. Hold times and recycling times are validated by applying the loads expected at each FPT stage after integration of the entire focal plane and optics tube using heaters. Figure~\ref{fig:frdige_cycle_hold_time_check} shows a typical millikelvin fridge cycle in a later run of the PB-2b backend with $t_{cycle} = 6$~hrs and $t_{ht} > 68$~hrs which demonstrates that $t_{tot}$ falls within a three sidereal day observation cycle. The loading for this cycle is less than the full loading after integration however, we still expect to achieve 72 hrs. Millikelvin validation of PB-2c is currently underway and will complete in the summer of 2018.

Additionally we obtain a measurement of the conductance of the millikelvin heatstraps, summarized in Table~\ref{tab:base_temps}, by observing the temperature gradient from the FPT stages to the He-10 heads and inferring the heat load from the He-10 load curves (Figure~\ref{fig:all_gl_plots}). Due to the strong temperature dependence of the Kapitza resistance below 1~K \cite{keller1971thermal, swartz1989thermal}, these values are somewhat variable depending on how the joint is prepared and the force applied \cite{didschuns2004thermal, gmelin1999thermal}. This indicates that neither the cross-sectional area of the C101 foils, nor the bulk conductivity is limiting $G$, rather it is the thermal interface resistance. While these values do vary from cooldown to cooldown, the numbers reported here are representative of the distribution of values of observed.

\section{Conclusion}
The Simons Array is a next generation CMB polarization experiment consisting of three PB-2 telescopes and cryogenic receivers and will observe at 95~GHz, 150~GHz, 220~GHz, and 270~GHz. Each receiver consists of two meter-scale cryostats -- the backend and the optics tube -- each employing PT415 cryocoolers to reach approximate temperatures of 50~K and 4~K. The backends use three-stage helium adsorption refrigerators to provide a 250~mK TES bolometer stage. In this work we have discussed the design, construction, and cryogenic validation of the PB-2b and -2c backend cryostats, which is focused on the desire for a three sidereal day observation cycle. In order to realize this goal, for PB-2b we have extensively characterized an \emph{off-the-shelf} He-10 fridge and modeled its condensation efficiency. This, along with thermal conductivity measurements of materials contributing loading to the millikelvin stages, motivates the design of this structure so we may lower loads and enable the fridge to achieve a three day cycle. We find that for our requriements the standard model He-10 gas light fridge from Chase Research Cryogenics has an undercharged He-4 stage which requires addition of the standalone booster He-4 fridge to meet this goal. For PB-2c we  purchased an overcharged super gas light He-10 which meets the three day hold time requirement in the dark configuration with no attached mass. As of spring 2018, the PB-2b backend is undergoing detector integration and readout commissioning before integration with its optics tube this summer-fall. Concurrently, the PB-2c backend is undergoing millikelvin cryogenic validation, after which detector and readout integration may begin.

\section{Acknowledgements}
POLARBEAR and The Simons Array are funded by the Simons Foundation and by grants from the National Science Foundation AST-0618398 and AST-1212230. All detector arrays for Simons Array are fabricated at the UC Berkeley Marvell Nanofabrication Laboratory. All silicon lenslet arrays are fabricated at the Nano3 Microfabrication Laboratory at UCSD. The Simons Array will operate at the James Ax Observatory in the Parque Astronomico Atacama in Northern Chile under the stewardship of the Comisi\'on Nacional de Investigacion Cientifica y Tecnol\'ogica de Chile (CONICYT).

\bibliography{references} 

\begin{thebibliography}{10}

\bibitem{ade2014detection}
P.~A.~R. Ade, R.~W. Aikin, D.~Barkats, S.~J. Benton, C.~A. Bischoff, J.~J.
  Bock, J.~A. Brevik, I.~Buder, E.~Bullock, C.~D. Dowell, L.~Duband, J.~P.
  Filippini, S.~Fliescher, S.~R. Golwala, M.~Halpern, M.~Hasselfield, S.~R.
  Hildebrandt, G.~C. Hilton, V.~V. Hristov, K.~D. Irwin, K.~S. Karkare, J.~P.
  Kaufman, B.~G. Keating, S.~A. Kernasovskiy, J.~M. Kovac, C.~L. Kuo, E.~M.
  Leitch, M.~Lueker, P.~Mason, C.~B. Netterfield, H.~T. Nguyen, R.~O'Brient,
  R.~W. Ogburn, A.~Orlando, C.~Pryke, C.~D. Reintsema, S.~Richter, R.~Schwarz,
  C.~D. Sheehy, Z.~K. Staniszewski, R.~V. Sudiwala, G.~P. Teply, J.~E. Tolan,
  A.~D. Turner, A.~G. Vieregg, C.~L. Wong, and K.~W. Yoon, ``{Detection of
  $B$-Mode Polarization at Degree Angular Scales by BICEP2},'' {\em Phys. Rev.
  Lett.}~{\bf 112}, p.~241101, Jun 2014.

\bibitem{mortonson2014joint}
M.~J. Mortonson and U.~Seljak, ``A joint analysis of planck and bicep2 b modes
  including dust polarization uncertainty,'' {\em Journal of Cosmology and
  Astroparticle Physics}~{\bf 2014}(10), p.~035, 2014.

\bibitem{arnold2014simons}
K.~Arnold, N.~Stebor, P.~Ade, Y.~Akiba, A.~E. Anthony, M.~Atlas, D.~Barron,
  A.~Bender, D.~Boettger, J.~Borrill, S.~Chapman, Y.~Chinone, A.~Cukierman,
  M.~Dobbs, T.~Elleflot, J.~Errard, G.~Fabbian, C.~Feng, A.~Gilbert,
  N.~Goeckner-Wald, N.~W. Halverson, M.~Hasegawa, K.~Hattori, M.~Hazumi,
  W.~Holzapfel, Y.~Hori, Y.~Inoue, G.~C. Jaehnig, A.~H. Jaffe, N.~Katayama,
  B.~Keating, Z.~Kermish, R.~Keskitalo, T.~Kisner, M.~{Le Jeune}, A.~Lee,
  E.~Leitch, E.~Linder, F.~Matsuda, T.~Matsumura, X.~Meng, N.~J. Miller,
  H.~Morii, M.~J. Myers, M.~Navaroli, H.~Nishino, T.~Okamura, H.~Paar,
  J.~Peloton, D.~Poletti, C.~Raum, G.~Rebeiz, C.~L. Reichardt, P.~L. Richards,
  C.~Ross, K.~M. Rotermund, D.~E. Schenck, B.~D. Sherwin, I.~Shirley, M.~Sholl,
  P.~Siritanasak, G.~Smecher, B.~Steinbach, R.~Stompor, A.~Suzuki, J.~Suzuki,
  S.~Takada, S.~Takakura, T.~Tomaru, B.~Wilson, A.~Yadav, and O.~Zahn, ``{The
  Simons Array: expanding POLARBEAR to three multi-chroic telescopes},'' {\em
  SPIE Astronomical Telescopes + Instrumentation} (August 2014), p.~91531F,
  2014.

\bibitem{stebor2016simons}
N.~Stebor, P.~Ade, Y.~Akiba, C.~Aleman, K.~Arnold, C.~Baccigalupi, B.~Barch,
  D.~Barron, S.~Beckman, A.~Bender, D.~Boettger, J.~Borrill, S.~Chapman,
  Y.~Chinone, A.~Cukierman, T.~de~Haan, M.~Dobbs, A.~Ducout, R.~Dunner,
  T.~Elleflot, J.~Errard, G.~Fabbian, S.~Feeney, C.~Feng, T.~Fujino, G.~Fuller,
  A.~J. Gilbert, N.~Goeckner-Wald, J.~Groh, G.~Hall, N.~Halverson, T.~Hamada,
  M.~Hasegawa, K.~Hattori, M.~Hazumi, C.~Hill, W.~L. Holzapfel, Y.~Hori,
  L.~Howe, Y.~Inoue, F.~Irie, G.~Jaehnig, A.~Jaffe, O.~Jeong, N.~Katayama,
  J.~P. Kaufman, K.~Kazemzadeh, B.~G. Keating, Z.~Kermish, R.~Keskitalo,
  T.~Kisner, A.~Kusaka, M.~{Le Jeune}, A.~T. Lee, D.~Leon, E.~V. Linder,
  L.~Lowry, F.~Matsuda, T.~Matsumura, N.~Miller, J.~Montgomery, M.~Navaroli,
  H.~Nishino, H.~Paar, J.~Peloton, D.~Poletti, G.~Puglisi, C.~R. Raum, G.~M.
  Rebeiz, C.~L. Reichardt, P.~L. Richards, C.~Ross, K.~M. Rotermund, Y.~Segawa,
  B.~D. Sherwin, I.~Shirley, P.~Siritanasak, L.~Steinmetz, R.~Stompor,
  A.~Suzuki, O.~Tajima, S.~Takada, S.~Takatori, G.~P. Teply, A.~Tikhomirov,
  T.~Tomaru, B.~Westbrook, N.~Whitehorn, A.~Zahn, and O.~Zahn, ``{The Simons
  Array CMB polarization experiment},'' {\em SPIE Astronomical Telescopes +
  Instrumentation} (July), p.~99141H, 2016.

\bibitem{kermish2012polarbear}
Z.~D. Kermish, P.~Ade, A.~Anthony, K.~Arnold, D.~Barron, D.~Boettger,
  J.~Borrill, S.~Chapman, Y.~Chinone, M.~A. Dobbs, J.~Errard, G.~Fabbian,
  D.~Flanigan, G.~Fuller, A.~Ghribi, W.~Grainger, N.~Halverson, M.~Hasegawa,
  K.~Hattori, M.~Hazumi, W.~L. Holzapfel, J.~Howard, P.~Hyland, A.~Jaffe,
  B.~Keating, T.~Kisner, A.~T. Lee, M.~{Le Jeune}, E.~Linder, M.~Lungu,
  F.~Matsuda, T.~Matsumura, X.~Meng, N.~J. Miller, H.~Morii, S.~Moyerman, M.~J.
  Myers, H.~Nishino, H.~Paar, E.~Quealy, C.~L. Reichardt, P.~L. Richards,
  C.~Ross, A.~Shimizu, M.~Shimon, C.~Shimmin, M.~Sholl, P.~Siritanasak,
  H.~Spieler, N.~Stebor, B.~Steinbach, R.~Stompor, A.~Suzuki, T.~Tomaru,
  C.~Tucker, and O.~Zahn, ``{The POLARBEAR experiment},'' {\em SPIE
  Astronomical Telescopes + Instrumentation} (September), p.~84521C, 2012.

\bibitem{levi2013desi}
M.~Levi, C.~Bebek, T.~Beers, R.~Blum, R.~Cahn, D.~Eisenstein, B.~Flaugher,
  K.~Honscheid, R.~Kron, O.~Lahav, and the DESI~collaboration, ``{The DESI
  Experiment, a whitepaper for Snowmass 2013},'' {\em arXiv preprint
  arXiv:1308.0847} , 2013.

\bibitem{ade2006review}
P.~A. Ade, G.~Pisano, C.~Tucker, and S.~Weaver, ``A review of metal mesh
  filters,'' in {\em Millimeter and Submillimeter Detectors and Instrumentation
  for Astronomy III},   {\bf 6275}, p.~62750U, International Society for Optics
  and Photonics, 2006.

\bibitem{green2015cooling}
M.~A. Green, S.~S. Chouhan, C.~Wang, and A.~F. Zeller, ``Second stage cooling
  from a cryomech pt415 cooler at second stage temperatures up to 300 k with
  cooling on the first-stage from 0 to 250 w,'' {\em IOP Conference Series:
  Materials Science and Engineering}~{\bf 101}(1), p.~012002, 2015.

\bibitem{pobell2007matter}
F.~Pobell, {\em Matter and methods at low temperatures}, vol.~2, Springer,
  2007.

\bibitem{cheng1996high}
E.~S. Cheng, S.~S. Meyer, and L.~A. Page, ``A high capacity 0.23 k 3he
  refrigerator for balloon-borne payloads,'' {\em Review of scientific
  instruments}~{\bf 67}(11), pp.~4008--4016, 1996.

\bibitem{bender2014multiplexing}
A.~N.~Bender, J.-F. Cliche, T.~de~Haan, M.~A.~Dobbs, A.~Gilbert, J.~Montgomery,
  N.~Rowlands, G.~M.~Smecher, K.~Smith, and A.~Wilson, ``Digital frequency
  domain multiplexing readout electronics for the next generation of millimeter
  telescopes,'' {\em SPIE Astronomical Telescopes + Instrumentation}~{\bf
  9153}, 07 2014.

\bibitem{howell2010thermal}
J.~R. Howell, M.~P. Menguc, and R.~Siegel, {\em Thermal radiation heat
  transfer}, CRC press, 2010.

\bibitem{bapat1990performance}
S.~Bapat, K.~Narayankhedkar, and T.~Lukose, ``Performance prediction of
  multilayer insulation,'' {\em Cryogenics}~{\bf 30}(8), pp.~700--710, 1990.

\bibitem{keller1971thermal}
C.~Keller, ``Thermal performance of multilayer insulations final report,''
  1971.

\bibitem{RunyanJones}
M.~Runyan and W.~Jones, ``Thermal conductivity of thermally-isolating polymeric
  and composite structural support materials between 0.3 and 4k,'' {\em
  Cryogenics}~{\bf 48}, pp.~448--454, 07 2008.

\bibitem{Kellaris2014}
N.~Kellaris, M.~Daal, M.~Epland, M.~Pepin, O.~Kamaev, P.~Cushman, E.~Kramer,
  B.~Sadoulet, N.~Mirabolfathi, S.~Golwala, and M.~Runyan, ``Sub-kelvin thermal
  conductivity and radioactivity of some useful materials in low background
  cryogenic experiments,'' {\em Journal of Low Temperature Physics}~{\bf 176},
  pp.~201--208, Aug 2014.

\bibitem{Barucci2000}
M.~Barucci, E.~Gottardi, I.~Peroni, and V.~Guido, ``Low temperature thermal
  conductivity of kapton and upilex,'' {\em Cryogenics}~{\bf 40}, pp.~145--147,
  02 2000.

\bibitem{Olson1993}
J.~Olson, ``Thermal conductivity of some common cryostat materials between 0.05
  and 2 k,'' {\em Cryogenics}~{\bf 33}(7), pp.~729 -- 731, 1993.

\bibitem{woodcraft2005recommended}
A.~L. Woodcraft, ``Recommended values for the thermal conductivity of aluminium
  of different purities in the cryogenic to room temperature range, and a
  comparison with copper,'' {\em Cryogenics}~{\bf 45}(9), pp.~626--636, 2005.

\bibitem{krueger1972initial}
W.~H. Krueger and S.~Pollack, ``The initial oxidation of aluminum thin films at
  room temperature,'' {\em Surface Science}~{\bf 30}(2), pp.~263--279, 1972.

\bibitem{pollack1969kapitza}
G.~L. Pollack, ``Kapitza resistance,'' {\em Reviews of Modern Physics}~{\bf
  41}(1), p.~48, 1969.

\bibitem{berman1956some}
R.~Berman, ``Some experiments on thermal contact at low temperatures,'' {\em
  Journal of Applied Physics}~{\bf 27}(4), pp.~318--323, 1956.

\bibitem{salerno2003thermal}
L.~J. Salerno and P.~Kittel, ``{Thermal contact conductance},'' {\em Journal of
  Materials Processing Technology}~{\bf 135}(2-3), pp.~204--210, 2003.

\bibitem{swartz1989thermal}
E.~T. Swartz and R.~O. Pohl, ``Thermal boundary resistance,'' {\em Reviews of
  modern physics}~{\bf 61}(3), p.~605, 1989.

\bibitem{didschuns2004thermal}
I.~Didschuns, A.~Woodcraft, D.~Bintley, and P.~Hargrave, ``Thermal conductance
  measurements of bolted copper to copper joints at sub-kelvin temperatures,''
  {\em Cryogenics}~{\bf 44}(5), pp.~293--299, 2004.

\bibitem{gmelin1999thermal}
E.~Gmelin, M.~Asen-Palmer, M.~Reuther, and R.~Villar, ``Thermal boundary
  resistance of mechanical contacts between solids at sub-ambient
  temperatures,'' {\em Journal of Physics D: Applied Physics}~{\bf 32}(6),
  p.~R19, 1999.

\end{thebibliography}
\bibliographystyle{spiebib} 

\end{document}